\documentclass[tightenlines,twoside,secnumarabic,
               onecolumn,floatfix,nofootinbib,showpacs,11pt]{revtex4}

\usepackage[english]{babel}
\usepackage{amsmath,amssymb}
\usepackage{graphicx}
\usepackage[sort&compress]{natbib}

\allowdisplaybreaks

\bibpunct{[}{]}{,}{n}{}{,}

\newcommand{\ev}[1]{\ensuremath{\left\langle #1 %
                     \right\rangle}} 

\newcommand{\kev}{{\rm keV}}

\newcommand{\gev}{{\rm GeV}}

\newcommand{\be}{\begin{equation}}
\newcommand{\ee}{\end{equation}}
\def\ltap{\ \raise.3ex\hbox{$<$\kern-.75em\lower1ex\hbox{$\sim$}}\ }
\def\gtap{\ \raise.3ex\hbox{$>$\kern-.75em\lower1ex\hbox{$\sim$}}\ }
\def\lsim{\ \raise.3ex\hbox{$<$\kern-.75em\lower1ex\hbox{$\sim$}}\ }
\def\gsim{\ \raise.3ex\hbox{$>$\kern-.75em\lower1ex\hbox{$\sim$}}\ }
\usepackage{pstricks}

\begin{document}

\title{LEP Shines Light on Dark Matter
}
\author{Patrick J.\ Fox$^1$}         \email[Email: ]{pjfox@fnal.gov}
\author{Roni Harnik$^1$}             \email[Email: ]{roni@fnal.gov}
\author{Joachim Kopp$^1$}            \email[Email: ]{jkopp@fnal.gov}
\author{Yuhsin Tsai$^{2,1}$}         \email[Email: ]{yt237@cornell.edu}
\affiliation{$^1$ Theoretical Physics Department \\
                  \mbox{Fermilab, P.O.~Box 500, Batavia, IL 60510, USA} \vspace{0.2cm}\\
             $^2$ Institute for High Energy Phenomenology \\
                  Newman Laboratory of Elementary Particle Physics \\
                  Cornell University, Ithaca, NY 14853, USA}
\date{\today} 

\begin{abstract}
Dark matter pair production at high energy colliders may leave observable signatures in the energy and momentum spectra of the objects recoiling against the dark matter.
We use LEP data on mono-photon events with large missing energy to constrain the
coupling of dark matter to electrons. 
Within a large class of models, our limits are complementary to and competitive with
limits on dark matter annihilation and on WIMP-nucleon scattering from indirect
and direct searches. Our limits, however, do not suffer from systematic and astrophysical  uncertainties associated with direct and indirect limits. 
For example, we are able to rule out light ($\lesssim 10$~GeV) thermal relic dark matter with universal couplings exclusively to charged leptons.  In addition, for dark matter mass below about 80~GeV, LEP limits are stronger than Fermi constraints on annihilation into charged leptons in dwarf spheroidal galaxies. Within its kinematic reach, LEP also provides the strongest constraints on the spin-dependent direct detection
cross section in models with universal  couplings to both quarks and
leptons. 
In such models the strongest limit is also set on spin independent scattering for  dark matter masses below $\sim 4\ \gev$.
Throughout our discussion, we consider both low energy effective theories of
dark matter, as well as several motivated renormalizable scenarios involving light mediators.
\end{abstract}

\begin{flushright}
FERMILAB-PUB-11-039-T
\end{flushright}

\maketitle
\newpage

\section{Introduction}

The search for dark matter and its interaction with standard model particles is actively pursued by experiments worldwide. Direct detection searches look for a feeble kick that a dark matter particle produces in recoiling off a nucleus. Indirect searches aim at the detection of the annihilation products of dark matter particles with each other in regions with a high density of dark matter. A signal in any experiment using either of these techniques requires the existence of a new interaction between dark matter and standard model particles. Direct and indirect searches, together with assumptions on the astrophysical dark matter density and velocity distributions, place bounds on such possible interactions.

The very same interactions may also lead to the production of dark matter at a high energy collider (with an appropriate beam of incoming particles). In this article we will explore possible couplings of dark matter to leptons and the limits on such couplings from the LEP experiments at CERN. There, the annihilation of an electron and a positron into an invisible dark matter pair may become visible if an additional hard photon is radiated during the collision, producing a distinct mono-photon signal. Since the LEP experiments did not observe an excess of mono-photon events beyond the expected background, a limit may be placed on the postulated interaction strength between dark matter and the standard model. These limits, in turn, can be reinterpreted as limits on both direct and indirect detection rates, independent of astrophysical and atomic uncertainties.

Previous work relating collider searches to direct and indirect searches for dark matter has focused on the Tevatron \cite{Bai:2010hh,Goodman:2010ku} and the LHC~\cite{Goodman:2010yf}. While these hadronic machines probe the dark matter couplings to light quarks, the LEP data we are going to study is sensitive to the dark matter-electron coupling. The potential limits from ILC mono-photons on a thermal relic that couples to leptons was studied in~\cite{Birkedal:2004xn}.  If dark matter were hadrophobic, as has been discussed~\cite{Bernabei:2007gr,  Fox:2008kb,Dedes:2009bk} (but disfavored~\cite{Kopp:2009et,Kopp:2010su}) as a possible explanation  of the DAMA~\cite{Bernabei:2008yi} and CoGeNT~\cite{Aalseth:2010vx} signals, as well as various cosmic ray anomalies, the LEP mono-photon searches would provide the only sensitive, model independent, collider limits for dark matter.  As we shall see, LEP searches can yield bounds on dark matter which are both competitive with, and complementary to, those placed by traditional dark matter searches.

The plan of this paper is as follows. In the next section we will introduce the effective theory formalism we will use in the first part of the paper. The list of operators we are going to consider will not be exhaustive, but will encompass the phenomenologically most relevant scenarios.  We include cases where the dark matter-lepton couplings are scalar, vector and axial-vector in nature\footnote{Throughout we consider the dark matter to be a Dirac fermion, since our bounds would not be altered significantly if dark matter is a Majorana fermion~\cite{Goodman:2010yf,Bai:2010hh,Goodman:2010ku}. We also do not consider scalar or vector dark matter, though we do not expect the limits to be qualitatively different.} which covers a broad range of phenomena, including spin independent and spin dependent scattering as well as annihilations which are either velocity suppressed or not.  In section~\ref{sec-limits} we will set limits on the various contact operators from the mono-photon search at LEP. Then, in sections~\ref{sec:dd} and~\ref{sec:id} we will translate our limits into bounds on dark matter nucleon scattering and dark matter self annihilation, respectively. We will compare our results to current direct and indirect searches. In section~\ref{sec:light} we will consider the possibility that the effective theory described in section~\ref{sec:ops} is not appropriate for calculating the production rate of dark matter pairs at LEP. We will discuss several renormalizable models in which a new gauge boson or a new scalar particle is introduced to mediate the interactions of dark matter with leptons. As we shall see, the inclusion of such particles can significantly alter LEP bounds, and in certain regimes the bounds become sensitive to the details of the UV completion. We will conclude in section~\ref{sec:conclusion}.

\section{The Interaction of Dark matter with Leptons}
\label{sec:ops}

In order to produce dark matter at LEP it must couple to electrons. In many models this may occur via the exchange of a heavy mediator that can be integrated out of the theory at low energies. In that case one can describe the phenomenology in an effective field theory with higher dimension operators coupling the dark matter particle $\chi$ to standard model leptons $\ell = e, \mu, \tau$. This allows us to consider a large variety of dark matter phenomena without committing to a particular high energy framework\footnote{Indeed, several recent studies have used effective theories to analyze and draw connections among dark matter 
experiments~\cite{Harnik:2008uu,Cao:2009uv,Cao:2009uw,Agrawal:2010fh,Fan:2010gt}.}. 
We will be considering the operators 
\begin{align}
  \mathcal{O}_V &= \frac{(\bar\chi\gamma_\mu\chi)(\bar \ell \gamma^\mu \ell)}{\Lambda^2} \,,
    & \text{(vector, $s$-channel)} \label{O1} \\
  \mathcal{O}_S &= \frac{(\bar\chi\chi)(\bar \ell \ell)}{\Lambda^2} \,,
    & \text{(scalar, $s$-channel)} \\
  \mathcal{O}_A &= \frac{(\bar\chi\gamma_\mu\gamma_5\chi)(\bar \ell \gamma^\mu\gamma_5 \ell)}{\Lambda^2} \,,
    & \text{(axial vector, $s$-channel)} \label{O2} \\
  \mathcal{O}_t &= \frac{(\bar\chi \ell)(\bar \ell \chi)}{\Lambda^2} \,,
    & \text{(scalar, $t$-channel)} \label{O3}
\end{align}
which capture the essential dark matter and collider phenomenology (e.g.\ spin dependent and spin independent scattering on nucleons as well as $s$- and $p$- wave annihilation). The classification of these operators as $s$-channel or $t$-channel refers to their possible UV-completion: \eqref{O1}--\eqref{O2} are most straightforwardly obtained in models in which dark matter is produced at LEP through a neutral $s$-channel mediator, while eq.~\eqref{O3} arises most naturally if the mediator is a charged scalar exchanged in the $t$-channel. With such a UV completion in mind, the suppression scale $\Lambda$ can be interpreted as the mass of the mediator $M$, divided by the geometric mean of its couplings to leptons, $g_\ell$, and dark matter, $g_\chi$: $\Lambda = M / \sqrt{g_\ell g_\chi}$. Note that we assume lepton flavor to be conserved in the dark matter interaction.  LEP can only constrain couplings to electrons, $\ell = e$, and in principle the suppression scale $\Lambda$ could be different for couplings to $\mu$ and $\tau$ leptons. In the following discussion, we will therefore consider both scenarios in which dark matter couples \emph{only} to electrons (i.e.\ $\Lambda = \infty$ for $\ell = \mu, \tau$) and scenarios in which dark matter couples in a flavor-universal way to all standard model leptons.  Note that the last operator, eq.~\eqref{O3}, may be transformed into a linear combination of the first three operators, plus pseudoscalar and tensor contributions, using the Fierz identities, but we include it separately here because it is a common outcome of supersymmetric theories. 

The effective theory described by equations~\eqref{O1}--\eqref{O3} is always a valid description of processes with low momentum transfer, in particular dark matter-nucleon scattering in direct detection experiments. In high energy processes such as dark matter production at LEP or dark matter annihilation, the effective theory breaks down if the 4-momentum transfer is comparable to or larger than the mass of the particle mediating the interaction. In the first part of our analysis in sections~\ref{sec-limits}--\ref{sec:id}, we assume that this is not the case, and derive bounds on the operators \eqref{O1}--\eqref{O3} from LEP mono-photon searches, which we will then translate into constraints on direct and indirect dark matter detection cross sections. In section \ref{sec:light} we will investigate how these bounds change if the mediator of dark matter interactions is light so that an effective theory description is no longer possible.

\section{LEP Limits on the effective Dark Matter--electron coupling}
\label{sec-limits}

\begin{figure}
  \begin{center}
    \includegraphics[width=8cm]{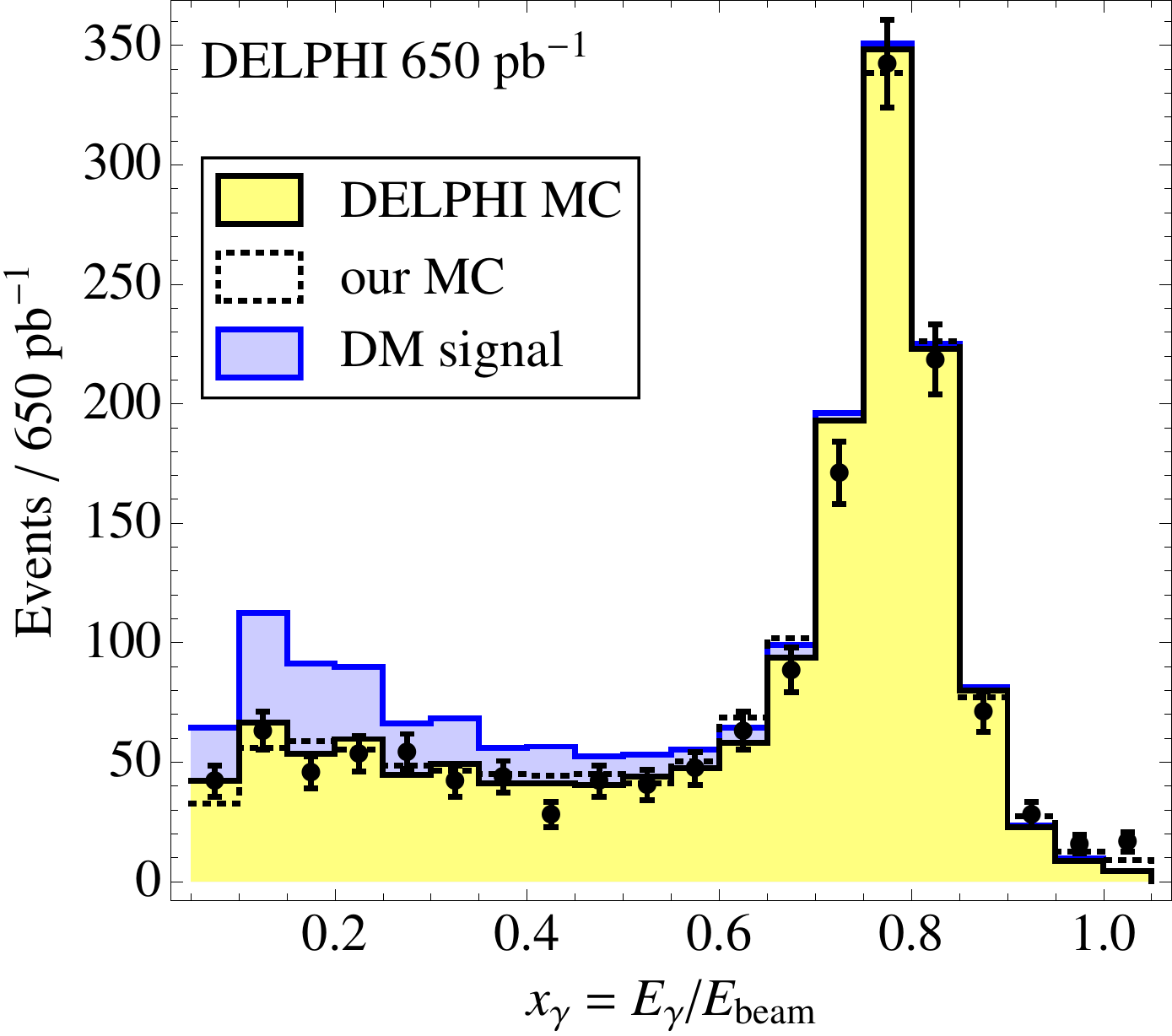}
  \end{center}
  \caption{Distribution of normalized photon energy in single-photon events at
    DELPHI. The agreement between the data (black dots with error bars) and both
    the full DELPHI Monte Carlo (solid yellow/light gray shaded histogram) as
    well as our CompHEP simulation (dotted histogram) is excellent. The blue
    shaded histogram shows what a hypothetical Dark Matter signal from $e^+ e^-
    \to \gamma\bar\chi\chi$ would look like. We have assumed vector-type contact
    interactions between electrons and dark matter, $m_\chi = 10$~GeV, and
    $\Lambda = 300$~GeV, see eq.~\eqref{O1}. The peak at $x_\gamma \sim 0.8$
    corresponds to the process $e^+ e^- \to \gamma Z^0 \to \gamma \nu \bar\nu$,
    with an on-shell $Z^0$.}
  \label{fig:data}
\end{figure}

In this section we will consider the operators ~\eqref{O1}--\eqref{O3} and
derive limits on their suppression scale $\Lambda$ from mono-photon searches at
LEP.  While all four LEP-detectors have studied single photon
events~\cite{LEPmonophoton}, we will here focus on data from the DELPHI
experiment~\cite{Abdallah:2003np,DELPHI:2008zg}, for which we were best able to
simulate the detector response.  The data was taken at center of mass
energies between 180~GeV and 209~GeV, but since in the analysis the events are
characterized only by the relative photon energy $x_\gamma = E_\gamma / E_{\rm
beam}$, we can make the simplifying assumption that all data was taken at an
energy of 100~GeV per beam. We have checked that the error introduced by this
approximation is small. For our Monte Carlo simulations, we use
CompHEP~\cite{Boos:2004kh,Pukhov:1999gg}, which allows us
to include the effect of initial state radiation (ISR) which we find to be
non-negligible. For example, we are only able to reproduce the height and width
of the on-shell $Z^0$ peak in the $x_\gamma$ distribution for the background
process $e^+ e^- \to \gamma \nu \bar\nu$ (cf.\ Figure~\ref{fig:data}) if ISR is
included.

To analyze the event samples generated in CompHEP, we use a modified version of
MadAnalysis~\cite{Alwall:2007st}, in which we have implemented the analysis
cuts and efficiencies of the DELPHI analysis as well as energy smearing
according to the resolution of the DELPHI electromagnetic calorimeters. In
doing so, we closely follow ref.~\cite{Abdallah:2003np}. 

In DELPHI, central photons with a polar angle $\theta$ (with respect to the
beam axis) in the range $45^\circ<\theta<135^\circ$ are detected in the High
Density Projection Chamber (HPC) with a threshold of $x_\gamma > 0.06$. We
assume the trigger efficiency for photons in the HPC to increase linearly from
52\% at $E_\gamma = 6$~GeV to 77\% at 30~GeV, and then to 84\% at 100~GeV.  The
trigger efficiency is multiplied by the efficiency of the subsequent analysis,
which we assume to increase linearly from 41\% at 6~GeV to 78\% at 80~GeV and
above. 

For photons with $12^\circ < \theta < 32^\circ$, detected in the Forward
Electromagnetic Calorimeter (FEMC), the threshold is $x_\gamma > 0.1$.  The
trigger efficiency increases linearly from 93\% at 10~GeV to 100\% at 15~GeV
and above, and the analysis efficiency is the product of a linear function,
increasing from 57\% at 10~GeV to 75\% at 100~GeV, and a constant 89\%, with
the first factor coming from the analysis cuts, and the second one describing
the loss of events due to noise and machine backgrounds. In addition, we impose
an energy dependent angular cut $\theta > (28 - 80 x_\gamma)^\circ$.  

Very forward photons ($3.8^\circ < \theta < 8^\circ$) give a signal in the
Small Angle Tile Calorimeter (STIC), whose threshold is $x_\gamma > 0.3$, and
we assume the efficiency to be 48\%, based on the (incomplete) information
given in~\cite{Abdallah:2003np}. We again impose an energy dependent angular
cut $\theta > (9.2 - 9 x_\gamma)^\circ$. 

The above, calorimeter specific, efficiencies are augmented by an additional
90\% efficiency factor, applied to all photons.  We found it necessary to
introduce this overall efficiency factor to gain agreement in normalization
between our simulations and the results of DELPHI.

The relative energy resolution, $\sigma_E/E$, is $0.043\,\oplus\,0.32/\sqrt{E}$
in the HPC, $0.03\,\oplus\,0.12/\sqrt{E}\,\oplus\,0.11/E)$ in the FEMC, and
$0.0152\,\oplus\,0.135/\sqrt{E}$ in the STIC, where $E$ is in units of GeV.
Here $\oplus$ means that the different contributions to the energy resolution
function are statistically independent.  For example, we simulate the effect of
finite energy resolution in the HPC by shifting the energy of each HPC photon
by an amount $0.043\,E \cdot r_1 + 0.32\,\sqrt{E} \cdot r_2$, where $r_1$ and
$r_2$ are independent Gaussian random numbers. Since we find that with purely
Gaussian energy smearing we are unable to reproduce the broad tails of the
on-shell $Z^0$ peak in the $x_\gamma$ distribution (Figure~\ref{fig:data}), we
impose an additional Lorentzian energy smearing with a width of $0.052\,E$. This
is motivated by a fit to the calorimeter response to monoenergetic electrons,
obtained from ref.~\cite{Feindt:2010}.

We have verified our modeling of the DELPHI detector by simulating the energy
distribution of single photons in the Standard Model. As demonstrated in
Figure~\ref{fig:data}, the agreement with the data (black dots with error bars)
and with the DELPHI Monte Carlo simulation (solid yellow/light gray histogram)
is excellent. Only in the very last bin ($x_\gamma > 1$), the observed number
of events is $\sim 4\sigma$ higher than the prediction by \emph{both} Monte
Carlo simulations, probably due to imperfect modeling of the detector
resolution function.  We therefore omit this bin in the following analysis. A
straightforward $\chi^2$ analysis then yields $\chi^2/{\rm dof} = 21.5/19$ for
our simulation, and $\chi^2/{\rm dof} = 20.6/19$ for the DELPHI Monte Carlo.

When setting limits on dark matter properties, we use our own simulation only
for the signal contribution, while the predicted backgrounds are taken from the
DELPHI Monte Carlo.  The blue shaded histogram in Figure~\ref{fig:data} shows what a
typical dark matter signal would look like for the case of operator
$\mathcal{O}_V$, with a dark matter mass of 10~GeV and with $\Lambda=300$~GeV.
Since most of the signal events are in the low-$x_\gamma$ region, where SM
backgrounds are only moderate, and since the spectral shape of the signal is
different from that of the background, we expect good sensitivity to the dark
matter-electron coupling $\Lambda^{-1}$.
\begin{figure}
  \begin{center}
    \includegraphics[width=8cm]{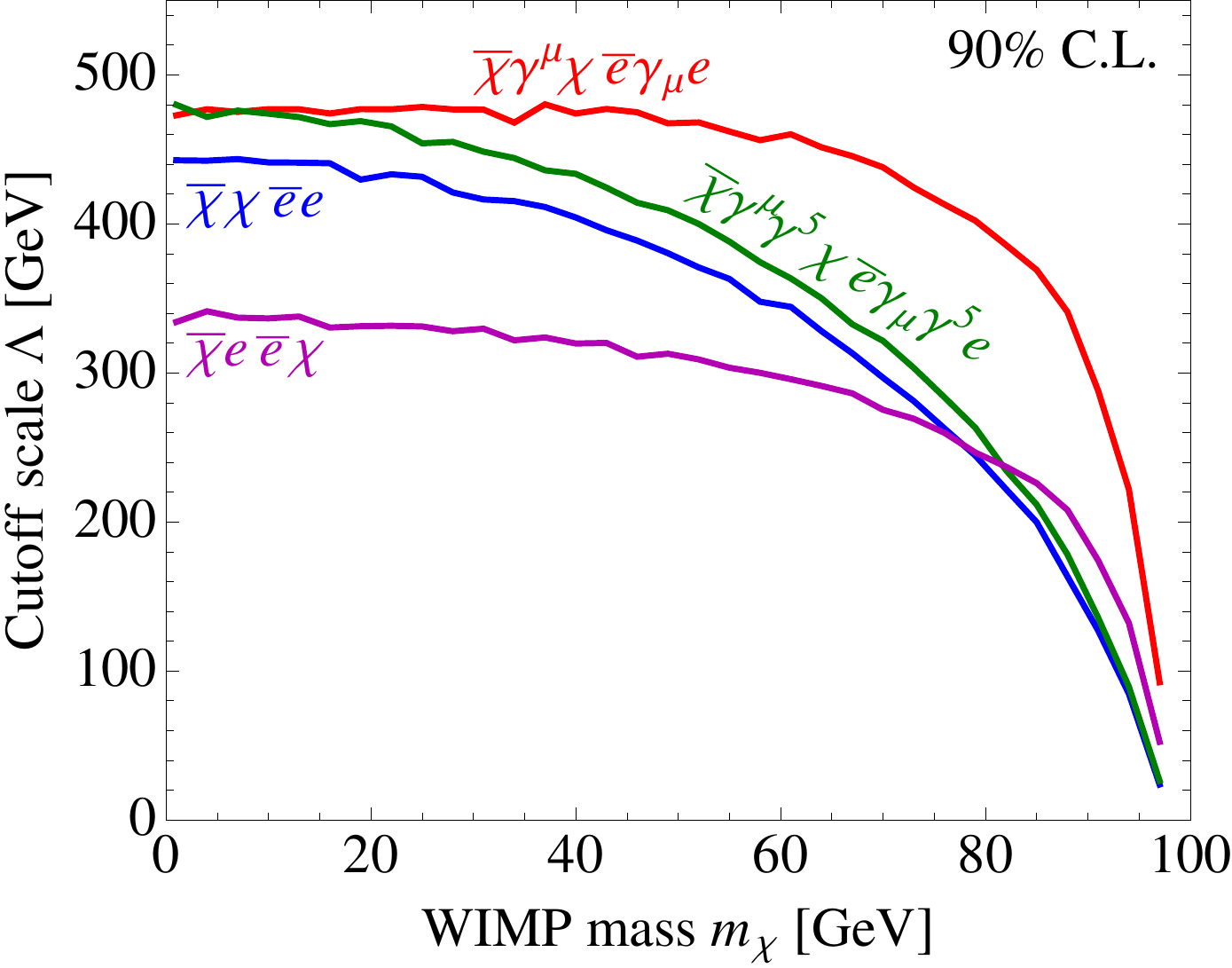}
  \end{center}
  \caption{DELPHI lower limits on the cutoff scale $\Lambda$ of
    the dark matter effective theory for the four operators
    eqs.~\eqref{O1}--\eqref{O3} as a function of the dark matter mass. The
    wiggles in the plot are due to limited Monte Carlo statistics.}
  \label{fig:cutoff}
\end{figure}

Indeed, a $\chi^2$ analysis yields limits on the cutoff scale $\Lambda$ of
order 250--500~GeV for dark matter masses $m_\chi \lesssim 80$~GeV (see
Figure~\ref{fig:cutoff}).  In this mass range, our limits on dark
matter-electron coupling are slightly better than the limits on dark
matter-quark couplings derived from Tevatron mono-jet
events~\cite{Goodman:2010ku,Bai:2010hh}. The Tevatron limits, however, do not
yet include spectral information, and they extend to dark matter masses of
several hundred GeV, while LEP is completely insensitive to $m_\chi \gtrsim
90$~GeV for kinematic reasons. The normalized photon energy distribution is
similar in shape for all the operators considered.  This leads to similar
limits on the operators from eqs.~\eqref{O1}--\eqref{O2} at low dark matter
mass. Only the limit on the strength of the operator $(\bar\chi \ell) (\bar
\ell \chi)$, eq.~\eqref{O3}, is somewhat weaker.  Using the Fierz identities
this operator may be converted to a sum of other operators involving a product
of a dark matter bilinear and a lepton bilinear.  There is destructive
interference between these operators leading to a smaller production cross
section for mono-photon events and thus a weaker bound on the cutoff scale for
this operator. When the dark matter mass $m_\chi$ exceeds $\sim 30$ GeV, the
limits on different operators scale differently with $m_\chi$ since at this
point the dark matter particles are produced closer to threshold and the
detailed dependence of the cross section on the final state velocities becomes
important.

\section{Limits on the Dark Matter--nucleon scattering cross section}
\label{sec:dd}

The next step is to translate the limits on $\Lambda$ into constraints on the
dark matter-nucleon scattering cross sections probed in direct detection experiments.
Since LEP can only probe dark matter-electron couplings, while direct detection
experiments are most sensitive to dark matter-quark couplings, this translation cannot
be done in a completely model-independent way. 
We thus consider two extreme possibilities, one in which the dark matter couples with equal strength to quarks as it does to leptons, and another in which dark matter couples only to leptons without coupling to quarks at tree level. Limits on other models, in which the ratio of lepton and quark couplings is different (e.g. coupling proportional to $B-L$),
may be easily derived from these two cases, as we shall see below. 

In order to compute the dark matter scattering cross section off a nucleon, $N = p, n$, through one of the operators in \eqref{O1}--\eqref{O3}, we need knowledge of the nucleon matrix elements $\langle N | \mathcal{O} | N \rangle$.  We use the values of these matrix elements presented in \cite{Bai:2010hh}, with the exception of $\langle N | \bar{q} q| N\rangle$ in which we follow \cite{Ellis:2008hf} but use an updated \cite{Young:2009ps} value of the pion-nucleon sigma term $\Sigma_{\pi N}=55$ MeV.~\footnote{Note however that recent lattice determinations~\cite{Young:2009zb,Toussaint:2009pz,Giedt:2009mr,:2010cw} of the strange quark content of the nucleus are considerably lower.  The effect on our bounds, assuming equal coupling to all fermions, is small.}  As mentioned earlier $\mathcal{O}_t$ can be converted from a ``$t$-channel'' operator to a sum of ``$s$-channel'' operators by use of Fierz identities.  Due to the relative size of the nucleon matrix elements it is sufficient to keep only the scalar $s$-channel contribution, which has a coefficient $1/4$.  Thus, for equal cutoff scale $\Lambda$, the direct detection rate expected from the operator $\mathcal{O}_t$ is the same as that expected from $\mathcal{O}_S/4$.

\begin{figure}
  \begin{center}
    \includegraphics[width=0.45\textwidth]{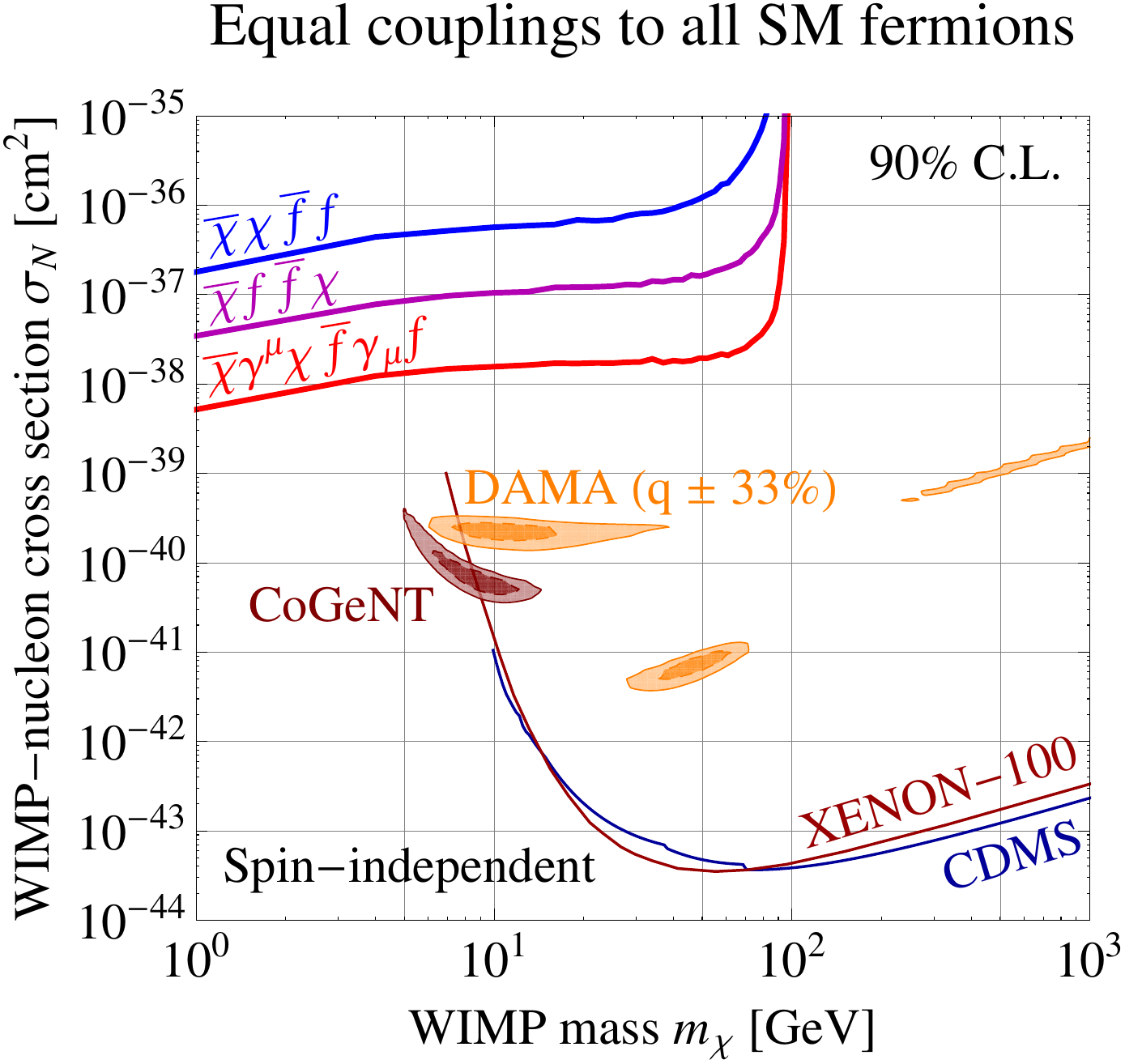}
    \includegraphics[width=0.45\textwidth]{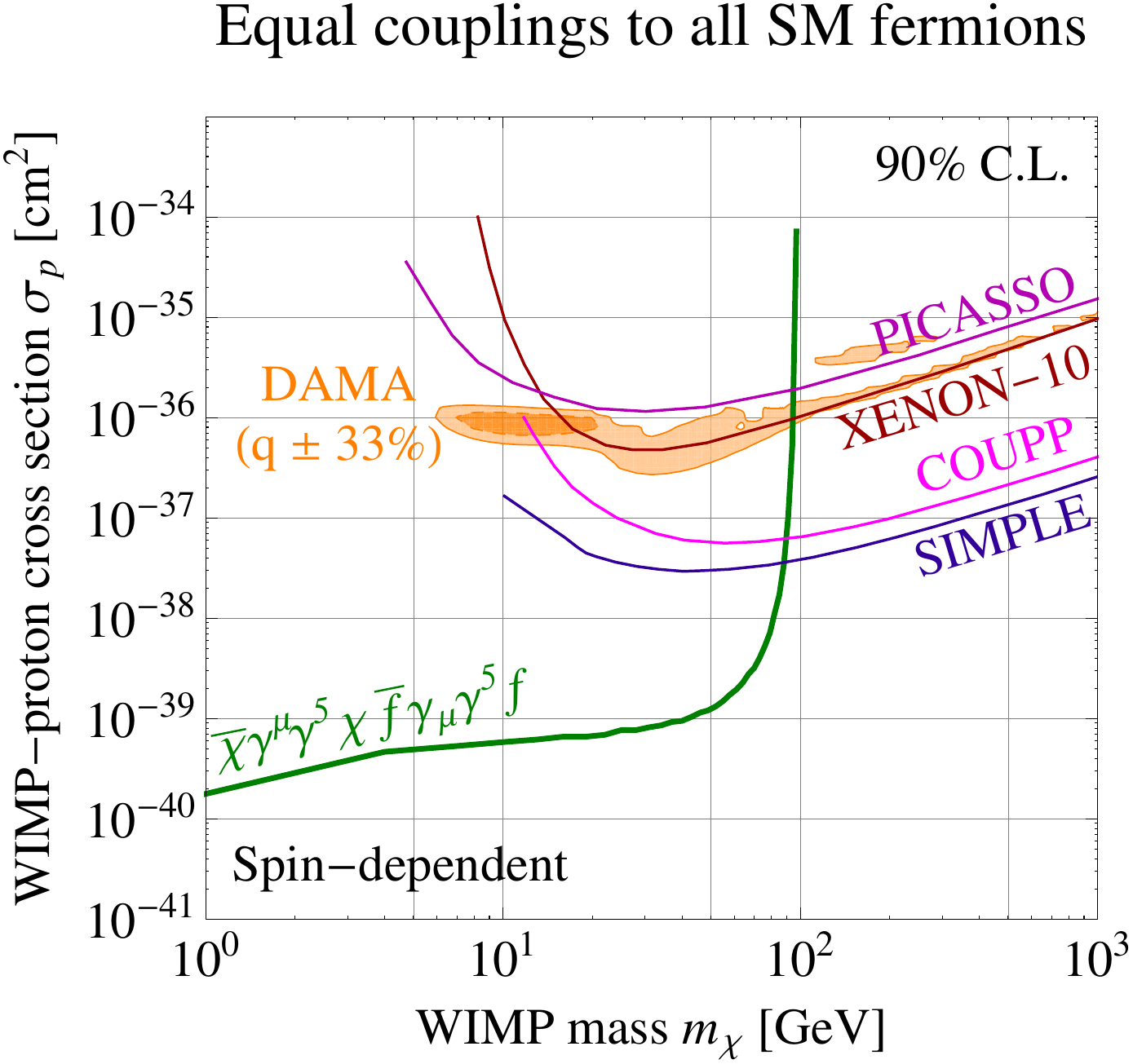}
  \end{center}
  \caption{DELPHI upper limits (thick lines) on the cross section for dark matter-nucleon
    scattering compared to results from direct detection experiments (thin
    lines and shaded regions).  The left-hand plot is for spin-independent
    scattering, as would come from operators $\mathcal{O}_S$, $\mathcal{O}_V$,
    $\mathcal{O}_t$, and the right is for spin-dependent scattering through
    operator $\mathcal{O}_A$.  The spin-independent limits of CDMS and XENON-100
    are taken from Refs.~\cite{Ahmed:2009zw} and \cite{Aprile:2010um},
    respectively. The spin-dependent limits of DAMA, XENON-10, PICASSO, COUPP
    and SIMPLE are taken from Refs.~\cite{Bernabei:2008yi},
    \cite{Angle:2008we}, \cite{BarnabeHeider:2005pg}, \cite{Behnke:2010xt}  and
    \cite{Girard:2011xc}, respectively.  The DAMA and CoGeNT-allowed regions
    are based on our own fit~\cite{Kopp:2009qt} to the data from
    Refs.~\cite{Bernabei:2008yi} and \cite{Aalseth:2010vx}. Following
    \cite{Hooper:2010uy}, we have conservatively assumed large systematic
    uncertainties on the DAMA quenching factors: $q_{\rm Na} = 0.3 \pm 0.1$ for
    sodium and $q_{\rm I} = 0.09 \pm 0.03$ for iodine. All limits are computed
    at the 90\% confidence level, while the DAMA and CoGeNT allowed regions
    are shown at the 90\% and $3\sigma$ confidence levels.}
  \label{fig:quarkcoupled}
\end{figure}

First we assume that the coupling of dark matter to all SM fermions, and in particular to all flavors of quarks, is identical to its couplings to electrons.  In this case, the LEP bound on $\Lambda$ can be immediately converted into an upper bound on the rate expected at direct detection experiments.  We show these bounds in Figure~\ref{fig:quarkcoupled} and we see that the
limits on spin-independent WIMP-nucleon scattering (left-hand plot) are
competitive with direct detection results only for very light dark matter, $m_\chi\lesssim 4\ \gev$.  The direct detection experiments become insensitive to such light masses due to their energy threshold, whereas there is no such low mass threshold at LEP.  The high mass cutoff at LEP is reflected in the rapid deterioration of the upper bound at $m_\chi \sim 90\ \gev$.  The LEP bound also applies directly to inelastic dark matter~\cite{TuckerSmith:2001hy}, since the splitting between the two dark matter states of $\sim 100\ \kev$ is inconsequential to the kinematics at LEP.  However, such models typically require considerably larger dark matter-nucleon cross sections than elastic dark matter, since the splitting allows only the high velocity fraction of the dark matter to scatter.  Our bounds derived from LEP rule out the very highest scattering cross sections in the parameter space consistent with DAMA~\cite{Kopp:2009qt}, but still leave the bulk of the parameter space allowed.

For spin-dependent scattering we expect the LEP bounds to be more competitive since there is little variation in the bound on $\Lambda$ between the operators responsible for spin-independent scattering ($\mathcal{O}_V$ and $\mathcal{O}_S$) and spin-dependent scattering ($\mathcal{O}_A$), whereas constraints from direct detection experiments are much weaker than in the spin-independent case. The reason for this is that, unlike spin-independent dark matter-nucleus scattering, spin-dependent scattering is not enhanced by a factor $A^2$, where $A$ is the nuclear mass number.  These considerations are reflected in the right-hand plot of Figure~\ref{fig:quarkcoupled} where the LEP limits surpass direct detection constraints for $m_\chi\lesssim 80\ \gev$ at which point the phase space for dark matter production at LEP again starts to shrink.

\begin{figure}[t]
  \begin{center}
    \includegraphics[width=6cm]{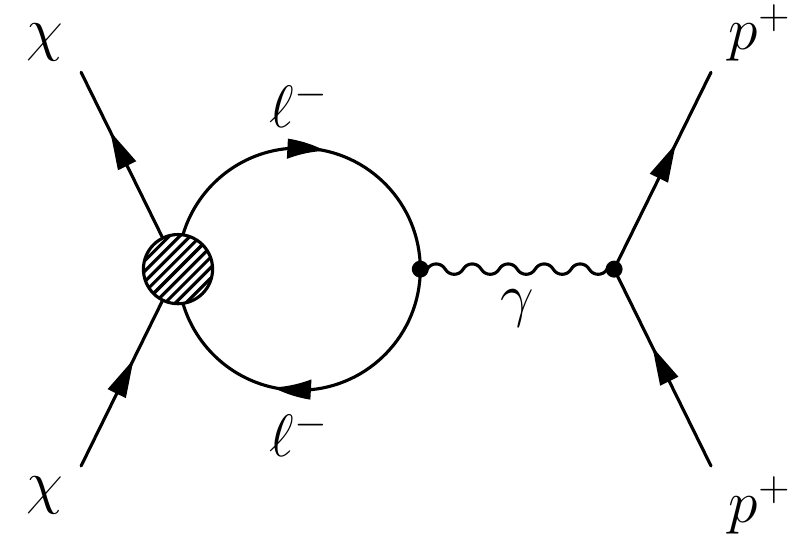}
  \end{center}
  \caption{Diagram for vector-type dark matter-proton scattering at the one-loop level.}
  \label{fig:1loop}
\end{figure}

If dark matter does not couple to quarks at tree level, but only to leptons
(for simplicity we assume the coupling to $\mu$ and $\tau$ is the same as that
to $e$, our conclusions are not significantly altered even if the coupling were
only to electrons), the power of the LEP limits improves dramatically.  The
reason is that in this case, dark matter-quark scattering to which direct
detection experiments are sensitive is only induced at the
loop-level~\cite{Kopp:2009et}.\footnote{Dark matter-electron scattering is
irrelevant in all direct detection experiments including
DAMA~\cite{Kopp:2009et} and CoGeNT~\cite{Kopp:2010su}. Even though DAMA and
CoGeNT would not reject bulk electron recoils as background, kinematics
dictates that the recoil energy can only be above the detection threshold if
the electron enters the interaction with an initial state momentum $\gtrsim 10$
MeV.  The probability for this is very small due to the fast drop-off of the
electron wave functions at high
momentum~\cite{Bernabei:2007gr,Kopp:2009et,Kopp:2010su}.} The cross section for
loop-induced dark matter-proton scattering through the diagram shown in
Figure~\ref{fig:1loop} is
\begin{align}
  \sigma_{\rm 1-loop} \simeq \frac{4 \alpha^2 \mu_p^2}{18^2 \pi^3 \Lambda^4} \cdot
    \Big[\!\! \sum_{\ell = e,\mu,\tau}f(q^2, m_\ell) \Big]^2 \,,
  \label{eq:1loop}
\end{align}
where $\alpha$ is the electromagnetic fine structure constant, $\mu_p = m_p m_\chi
/ (m_p + m_\chi)$ is the dark matter-proton reduced mass, and the loop factor
$f(q^2,m_\ell)$ is given by
\begin{align}
  f(q^2, m_\ell) = \frac{1}{q^2} \bigg[ 5 q^2 + 12 m_\ell^2 + 6 (q^2 + 2 m_\ell^2)
  \sqrt{1 - \frac{4 m_\ell^2}{q^2}} {\rm arcoth}\bigg(\sqrt{1 -  \frac{4 m_\ell^2}{q^2}}\bigg)
  - 3 q^2 \log\big[m_\ell^2 / \Lambda_{\rm ren}^2 \big] \bigg] \,.
  \label{eq:loop-factor}
\end{align}

We take the renormalization scale $\Lambda_{\rm ren}$ to be equal to $\Lambda$. Moreover, we make the approximation that all the dark matter is moving at the local escape velocity, which we take to be $v_\chi = 500$~km/sec, and that the momentum exchanged in the scattering is maximal, i.e.\ the scattering angle is $180^\circ$ in the center-of-momentum frame.  This will overestimate the rate of observed recoils at a direct detection experiment and will lead to a conservative upper bound.  With these assumptions the four-momentum exchanged between the dark matter and the target nucleus is $q^2 = - 4 \mu^2 v_\chi^2$, where $\mu$ is the invariant mass of the dark matter particle and the target nucleus.

The bounds on dark matter-nucleon cross sections quoted by direct detection experiments are derived from the actually measured dark matter-nucleus cross sections under the assumption that the dark matter couples equally to protons and neutrons and that the cross section is independent of $q^2$. Here, however, it only couples to protons and there is a $q^2$ dependence in the loop factor $f(q^2, m_\ell)$.  Thus, to enable a straight comparison, we rescale the quoted bounds on $\sigma_p$ by $A^2/Z^2 \times (\sum_\ell f(q_p^2,m_\ell)/\sum_\ell f(q^2/m_\ell))^2$, with $q_p^2= -4 \mu_p^2 v_\chi^2$; and we take $\Lambda_{\rm ren} = 500\ \gev$, the result is only very weakly sensitive to this choice.  Note that \eqref{eq:1loop} and \eqref{eq:loop-factor} are only approximations in the effective theory formalism. The exact form of the loop factor depends on the embedding of the effective theory into a complete renormalizable model.

\begin{figure}[t]
  \begin{center}
    \includegraphics[width=8cm]{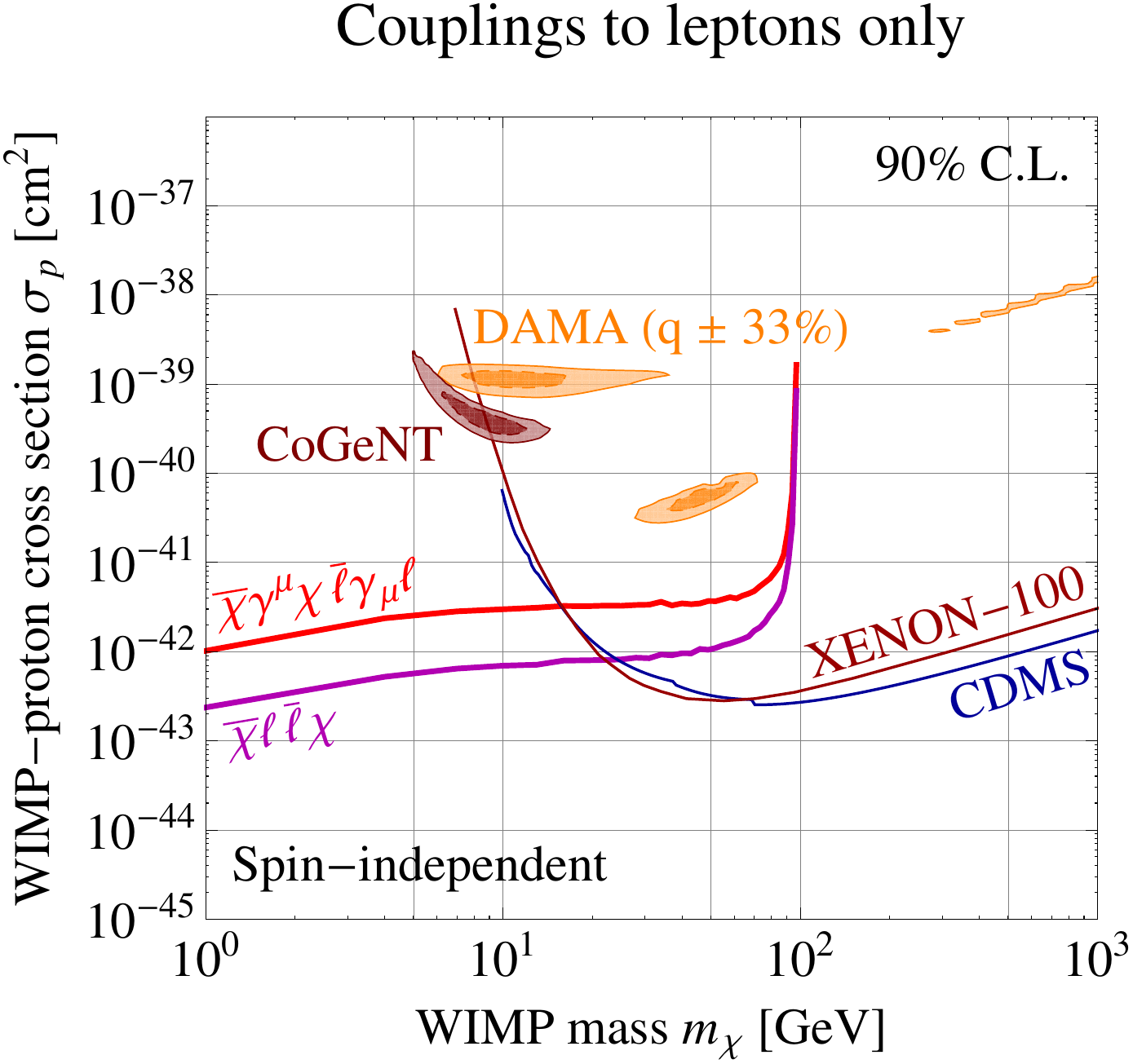}
  \end{center}
  \caption{DELPHI upper limits on the cross section for spin-independent dark matter--nucleon
    scattering for the case of dark matter with tree level couplings only to electrons,
    but loop level couplings also to quarks, compared to results from the direct detection
    experiments DAMA~\cite{Bernabei:2008yi}, CoGeNT~\cite{Aalseth:2010vx},
    CDMS~\cite{Ahmed:2009zw}, and XENON-100~\cite{Aprile:2010um}.  The DAMA and
    CoGeNT allowed regions are based on our own fit~\cite{Kopp:2009qt} to the
    data from refs.~\cite{Bernabei:2008yi, Aalseth:2010vx}. We conservatively
    assume $q_{\rm Na} = 0.3 \pm 0.1$ and $q_{\rm I} = 0.09 \pm 0.03$ for the
    DAMA quenching factors. All limits are computed at the 90\% confidence
    level, while the DAMA and CoGeNT allowed regions are shown at the 90\% and
    $3\sigma$ confidence levels.}
  \label{fig:leptononly}
\end{figure}

In Figure~\ref{fig:leptononly} we show the LEP bounds on dark matter in the
absence of tree-level couplings to quarks. Since loop-induced dark
matter-nucleon scattering is forbidden for axial-vector interactions and
suppressed by two loops for $s$-channel scalar interactions~\cite{Kopp:2009et},
we consider only the vector-type operator $\mathcal{O}_V$ and the scalar $t$-channel
operator $\mathcal{O}_t$. As before, we apply the Fierz identity to
$\mathcal{O}_t$ to decompose the operator into a linear combination of
$s$-channel operators, of which we keep only the vector contribution. As is
apparent from Figure~\ref{fig:leptononly}, an explanation of the DAMA and/or
CoGeNT signal by a dominantly leptophilic dark matter candidate which couples
to nuclei only through loops is ruled out by LEP.

Here we only considered two benchmark cases, where dark matter couples universally to SM fermions and when it couples only to leptons. Constraining a more general theory with a particular ratio of quark to lepton couplings, $R_{q/l}$, is straightforward. In this more general case nuclear recoil proceeds via both mechanisms, direct couplings to quarks and via a lepton loop. The limit on the former may be obtained by rescaling the bounds of Figure~\ref{fig:quarkcoupled} by $R_{q/l}^2$, whereas the limit on the latter may be taken directly from Figure~\ref{fig:leptononly}. Generically one of these limits will dominate the other over the full dark matter mass range, and the less constraining bound should be taken.

\section{Limits on the Dark Matter annihilation cross section}
\label{sec:id}

%
\begin{figure}[ht]
  \begin{center}
    \includegraphics[width=\textwidth]{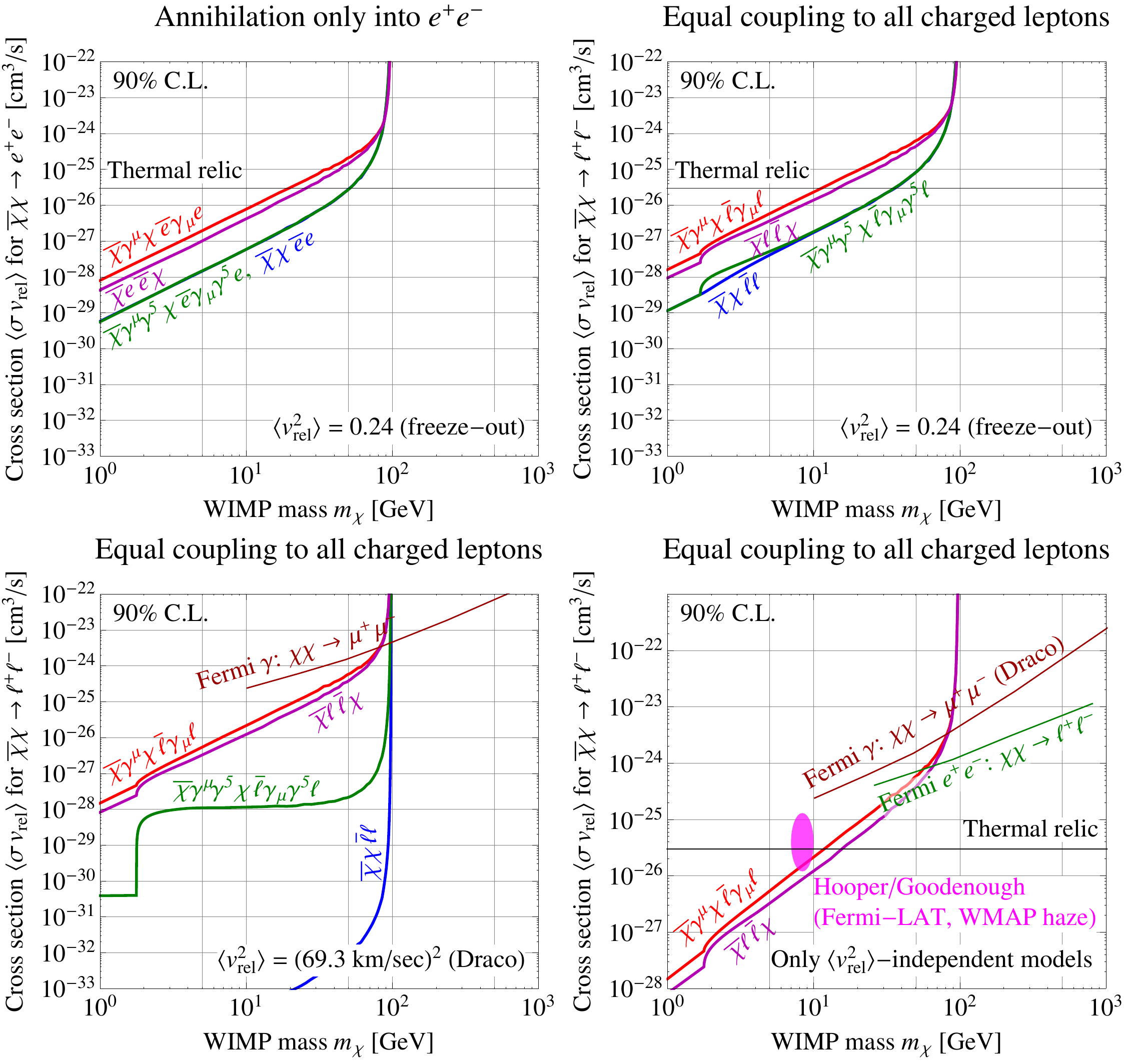}
  \end{center}
  \caption{LEP upper limits on the dark matter annihilation cross section
    $\ev{\sigma v}$, assuming that dark matter production at LEP and
    dark matter annihilation as probed by astrophysical and cosmological
    observations can be described by contact operators. In the upper left
    panel, we show limits on the process $\bar\chi \chi \to e^+ e^-$
    (the only one that can be constrained model-independently by LEP),
    while in the other panels we have made the assumption that dark matter
    couples equally to all charged leptons. For the average dark matter
    velocity $\ev{v^2}$
    we have
    assumed the value at freeze-out in the top panels, while the
    bottom left panel is for the Draco dwarf galaxy which has very
    small $\ev{v^2}$. In the bottom right panel we compare the LEP limit
    on the $v$-independent interactions, $\mathcal{O}_V$ and $\mathcal{O}_t$,
    to limits from a variety of astrophysical observations~\cite{Abdo:2010ex,
    Grasso:2009ma, Hooper:2010mq}.}
  \label{fig:annihilation}
\end{figure}

The LEP constraints on the suppression scale $\Lambda$ of the effective dark
matter couplings can also be converted to an upper bound on the annihilation
cross section of dark matter into an electron-positron pair. They can then be
compared to results from astrophysical probes of dark matter annihilation.
Moreover, if dark matter is a thermal relic and if annihilation into electrons
and positrons is the dominant annihilation channel, a lower bound on the dark
matter abundance in the universe can be derived. 
If dark matter has also other
annihilation modes, this bound is weakened by a factor $1 /
\text{BR}(\bar{\chi}\chi \to e^+ e^-)$.

In order to translate the LEP constraints on the coupling strength
$\Lambda^{-1}$ into limits on dark matter annihilation, we need to calculate
the annihilation cross sections corresponding to the operators in
equations~(\ref{O1})--(\ref{O3}).  For annihilation into a single single lepton
flavor of mass $m_\ell$, they read
\begin{align}
\sigma_S v_{\rm rel} & = \frac{1}{8\pi\Lambda^4}\sqrt{1-\frac{m_\ell^2}{m_\chi^2}}(m_\chi^2-m_\ell ^2)\, v_{\rm rel}^2 \,,
\label{eq:annrates1} \\
\sigma_V v_{\rm rel} & = \frac{1}{48\pi\Lambda^4}\sqrt{1-\frac{m_\ell ^2}{m_\chi^2}} \left(24 (2m_\chi^2+m_\ell ^2) + \frac{8 m_\chi^4-4m_\chi^2 m_\ell ^2+5m_\ell ^4}{m_\chi^2-m_\ell ^2}\,v_{\rm rel}^2\right), \\
\sigma_A v_{\rm rel} & = \frac{1}{48\pi\Lambda^4}\sqrt{1-\frac{m_\ell ^2}{m_\chi^2}}\left( 24  m_\ell ^2  +  \frac{8 m_\chi^4-22m_\chi^2 m_\ell ^2+17m_\ell ^4}{m_\chi^2-m_\ell ^2}\,v_{\rm rel}^2\right), \label{eq:annrates3} \\
\sigma_t v_{\rm rel} & = \frac{1}{192 \pi\Lambda^4}\sqrt{1-\frac{m_\ell ^2}{m_\chi^2}}
\left(24(m_\chi+m_\ell)^2 +   \frac{(m_\chi+m_\ell)^2(8 m_\chi^2-16 m_\chi m_\ell +11 m_\ell ^2)
}{m_\chi^2-m_\ell ^2}\,v_{\rm rel}^2 \right).
\label{eq:annrates4}
\end{align}
Here, we have made an expansion up to second order in the relative velocity
$v_{\rm rel}$ of the annihilating particles. While $v_{\rm rel} \ll 1$ in all
relevant astrophysical and cosmological environments, its exact value ranges
from $v_{\rm rel} \sim 0.1$ at the time of dark matter
decoupling in the early universe (if dark matter is a thermal relic) to values
of order $v_{\rm  rel}\ltap 10^{-4}$ (less than 100~km/s) in dwarf galaxies (see
Appendix~\ref{sec:vsq}).  This large spread of relative velocities can have a
large effect on annihilation rates for certain operators. Notably, annihilation
through an $s$-channel scalar operator, \eqref{eq:annrates1} is
suppressed by $v_{\rm rel}^2$, and annihilation through an $s$-channel axial
vector operator, \eqref{eq:annrates3} is suppressed by $v_{\rm rel}^2$
or by $m_\ell^2 / m_\chi^2$ compared to the other modes. The production cross
section at LEP is not suppressed in either of these cases, giving our bounds on
the suppressed modes a substantial relative advantage compared to indirect
searches. However, we will see that even in cases where the annihilation rate
is unsuppressed the LEP bounds are interesting and competitive for light dark
matter.

In Figure~\ref{fig:annihilation}, we consider both annihilation in the early
universe and annihilation in the Draco dwarf galaxy\footnote{We chose the Draco
dwarf galaxy because it is the dwarf galaxy for which Fermi-LAT obtains the
strongest bounds on dark matter annihilation~\cite{Abdo:2010ex}.} and compare
to the cross section required for a thermal relic ($\ev{\sigma v} \approx
3\times 10^{-26}\ \mathrm{cm}^3/\mathrm{s}$) and to several astrophysical
bounds.  Our most model-independent bounds, those on annihilation into $e^+e^-$,
are shown in the top left panel of Figure~\ref{fig:annihilation}, where we take
$\ev{v_{\rm rel}^2} = 0.24$, corresponding to thermal freeze-out. We see
that, if the dark matter only annihilates to electron-positron pairs, the
thermal relic cross section is ruled out by LEP at 90\% C.L.\ if $m_\chi
\lesssim 20$~GeV for vector interactions, and if $m_\chi \lesssim 50$~GeV for
scalar and axial vector interactions.  Thus, in order for such dark matter to be a thermal relic it must have additional annihilation modes.

Even though model-independently LEP can only constrain the dark matter coupling
to electrons and hence the annihilation cross section for the process $\bar\chi
\chi \to e^+ e^-$, in many models of dark matter the annihilation rate into
electrons is either equal or not very far from that into $\mu$ and $\tau$.  
For example, in models of supersymmetry the annihilation
rate into charged leptons is set by the slepton masses, which in many cases differ
by less than an order of magnitude. In
other models, such as universal extra dimensions the annihilation rates to
electrons, muons and taus are identical.\footnote{In models of universal extra
dimensions, dark matter is usually a vector particle, a case we are not considering in this work.} In order to present
our results we pick the simple benchmark in which the operator strengths are
universal among charged leptons. Constraints on other models may be derived from this
benchmark by the appropriate rescaling.  Limits on this benchmark scenario are
presented in the upper right, lower left, and lower right panels of
Figure~\ref{fig:annihilation}. Due to the strong dependence of $\ev{\sigma v}$
on the charged lepton mass for axial vector interactions, the limit on the
combined cross section for annihilation into all charged lepton species becomes
significantly stronger below the $\tau$ threshold in this case. In the lower
left panel of Figure~\ref{fig:annihilation}, we compare the LEP limits to
constraints from Fermi-LAT observations of the Draco dwarf galaxy in gamma
rays~\cite{Abdo:2010ex}.\footnote{The Fermi-LAT collaboration presented their
results as limits on the annihilation mode $\bar{\chi}\chi \to \mu^+ \mu^-$,
assuming that this is the only annihilation channel. We have reinterpreted
these limits, assuming that the branching ratio for the $\mu^+ \mu^-$ mode is
$1/3$ and that the $\gamma$-ray production is equal for all lepton flavors.  In reality this will not be true, in particular there will be additional hard photon production for the $\tau$ final state.   A reanalysis of Fermi-LAT data including gamma rays from annihilation channels other than $\mu^+ \mu^-$ could improve the limits by an $\mathcal{O}(1)$
factor.} For $m_\chi \lesssim 80$~GeV, LEP is superior to Fermi for all annihilation
operators considered here, especially for scalar interactions, for which
$\ev{\sigma v}$ is proportional to $\ev{v^2}$, which is extremely small in a
dwarf galaxy (see Appendix~\ref{sec:vsq}). 

In the lower right panel of Figure~\ref{fig:annihilation}, we have compiled
several constraints on dark matter annihilation in our galaxy. Since the dark
matter velocity distribution, especially at the galactic center is very
uncertain, we include only the predictions for annihilation through the operators $\mathcal{O}_V$
and $\mathcal{O}_t$ which is $\ev{v_{\rm rel}^2}$-independent and not
suppressed by the small lepton masses.  Limits on annihilation through the
axial vector operator $\mathcal{O}_A$ or the scalar operator $\mathcal{O}_S$
will be between the corresponding constraints at freeze-out and those from the
Draco dwarf galaxy and thus much stronger than the limits on vector
interactions. Comparing the LEP constraint to limits from astrophysical
observations, we find that the LEP limit is superior to Fermi results on gamma
rays from dwarf galaxies~\cite{Abdo:2010ex} and on the high energy $e^+
e^-$ spectrum~\cite{Grasso:2009ma}.  We also find that the excess in gamma rays at the galactic center
which has recently been argued~\cite{Hooper:2010im} to plausibly arise from
dark matter annihilations into $\tau$ leptons is also strongly constrained, \emph{if}
the annihilation proceeds into electron-positron pairs at a similar rate. In
fact, in~\cite{Hooper:2010im} it was argued that an equal annihilation rate
into electrons is favored because it may potentially provide an explanation of
the WMAP Haze~\cite{Finkbeiner:2003im,Finkbeiner:2004us}.  

Constraints on dark matter properties from both indirect and direct observations are sensitive to the abundance, and velocity distribution, of dark matter both locally, at the center of the galaxy and in sub-halos.  There are considerable uncertainties in all these quantities~\cite{Fox:2010bu} that effect the exclusion curves, and preferred regions in Figure~\ref{fig:annihilation}.  We emphasize that the LEP constraints do not suffer from these astrophysical uncertainties.

\section{Constraints on theories with light mediators}
\label{sec:light}

So far we have worked in a regime where the dark matter is the only particle of the dark sector accessible at colliders~\cite{Beltran:2010ww} and as a result all couplings of dark matter to the standard model are through higher dimension contact operators.  However, since LEP is a high energy machine, there is a possibility that the particle that is mediating the interaction of dark matter with electrons is light enough to cause significant deviations from the mono-photon rates and spectra predicted by the effective theory. These deviations will be most pronounced when the mediator is produced 
on-shell and then decays to a dark matter pair, but as we shall see, order one deviations are possible even without on-shell production. We therefore also consider LEP bounds for several renormalizable ``UV completions'' of our effective theory. 

\begin{figure}[ht]
  \begin{center}
    \includegraphics[width=8cm]{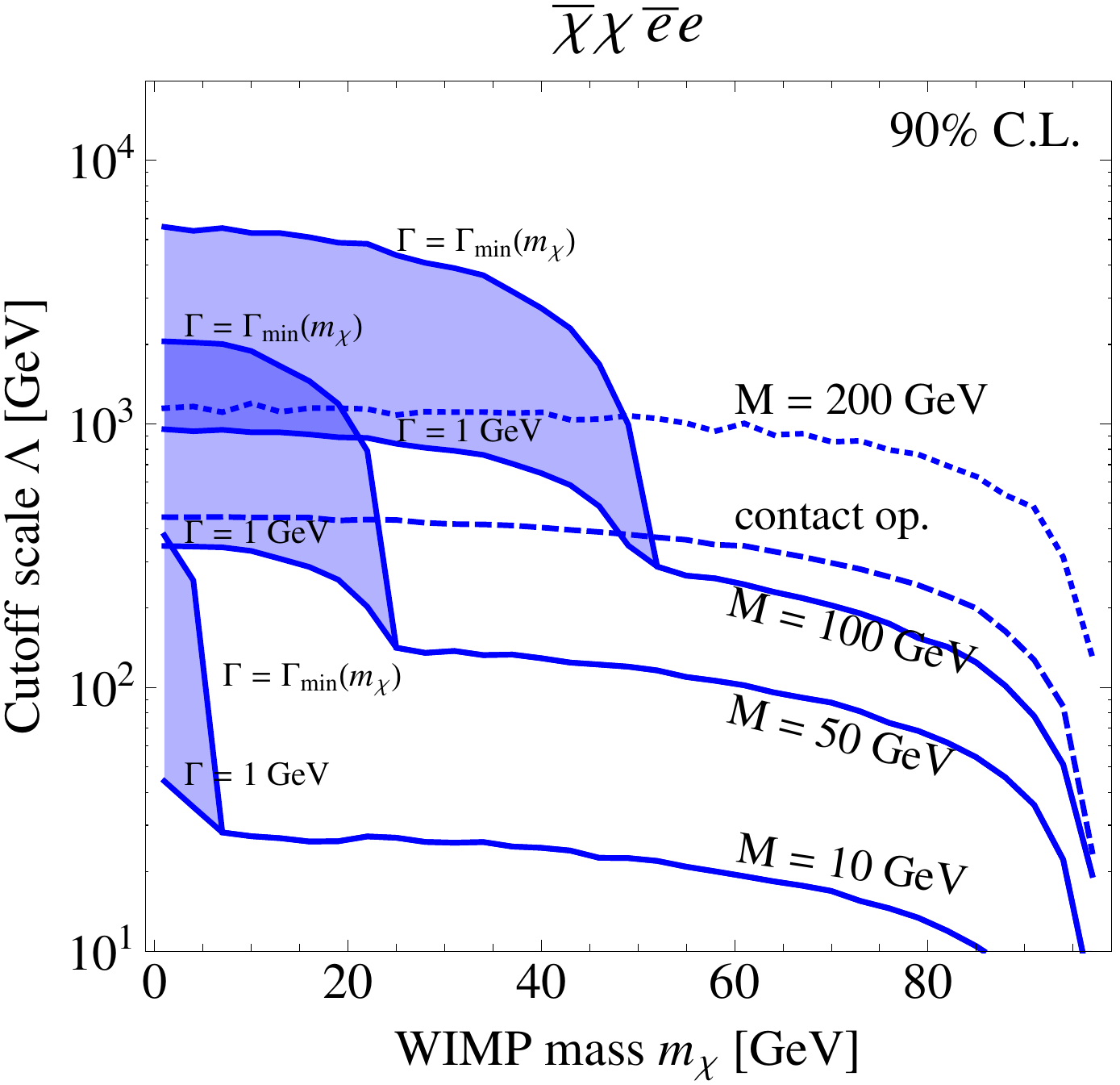}
    \includegraphics[width=8cm]{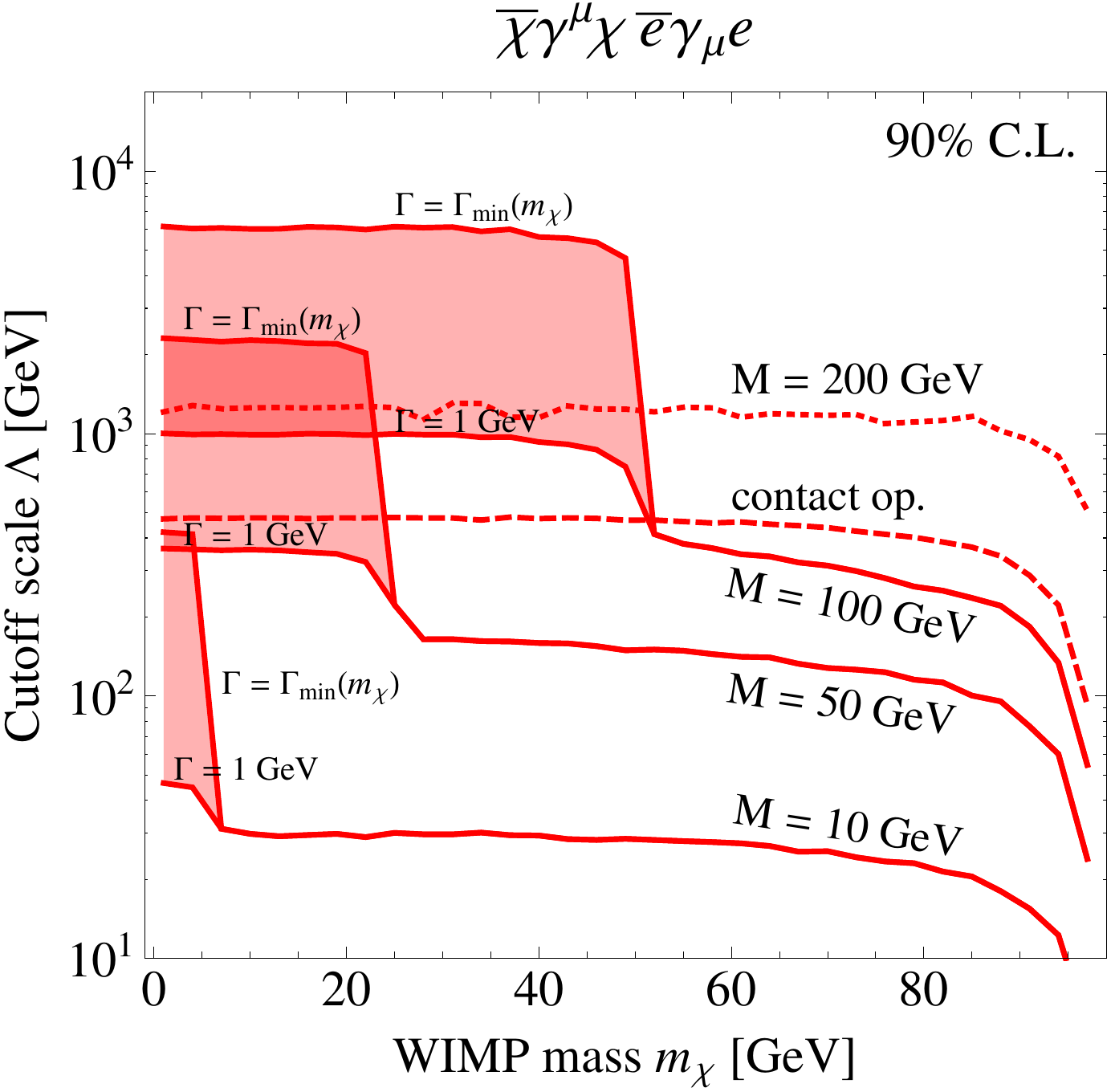} \\[0.2cm]
    \includegraphics[width=8cm]{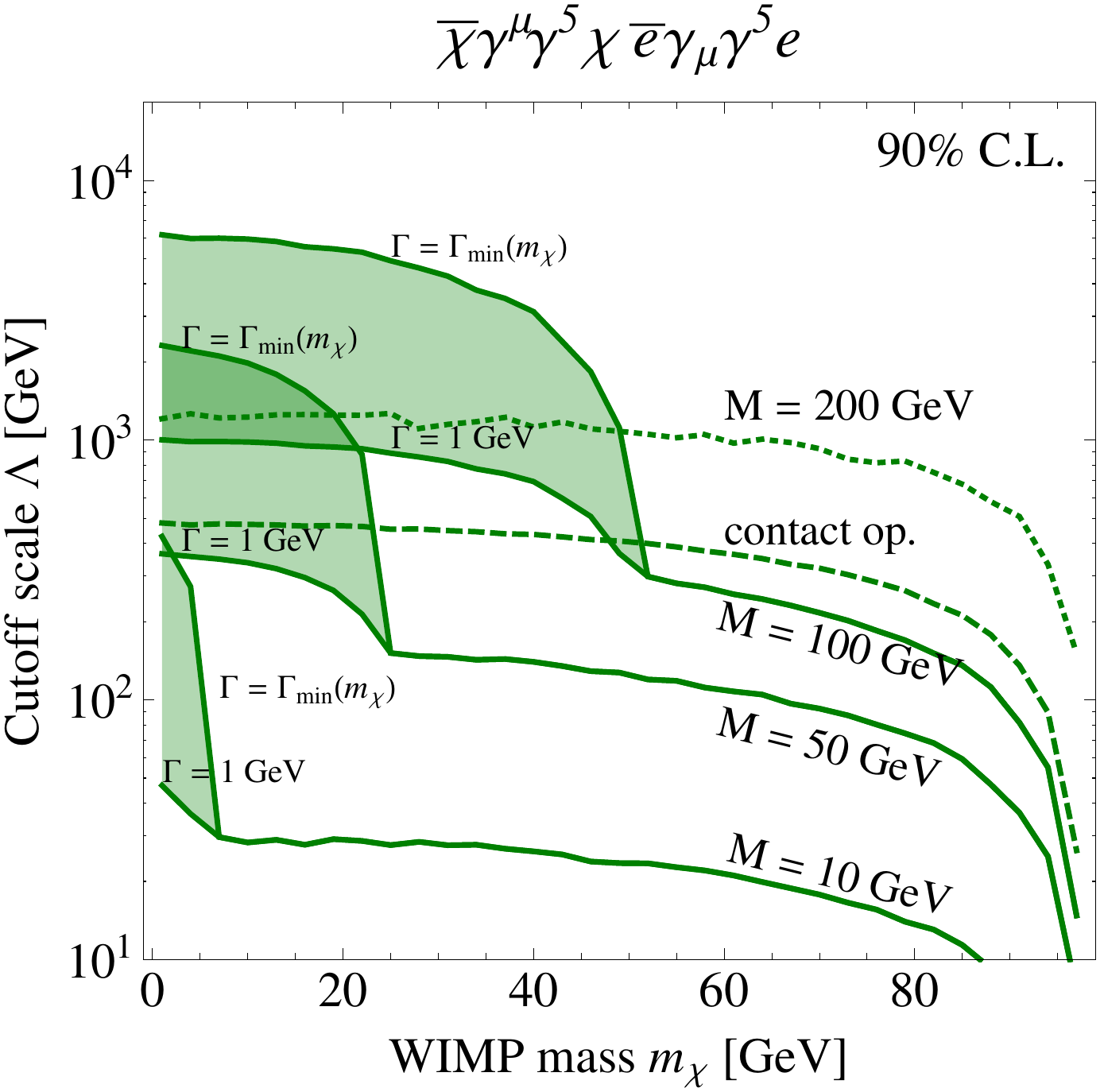}
    \includegraphics[width=8cm]{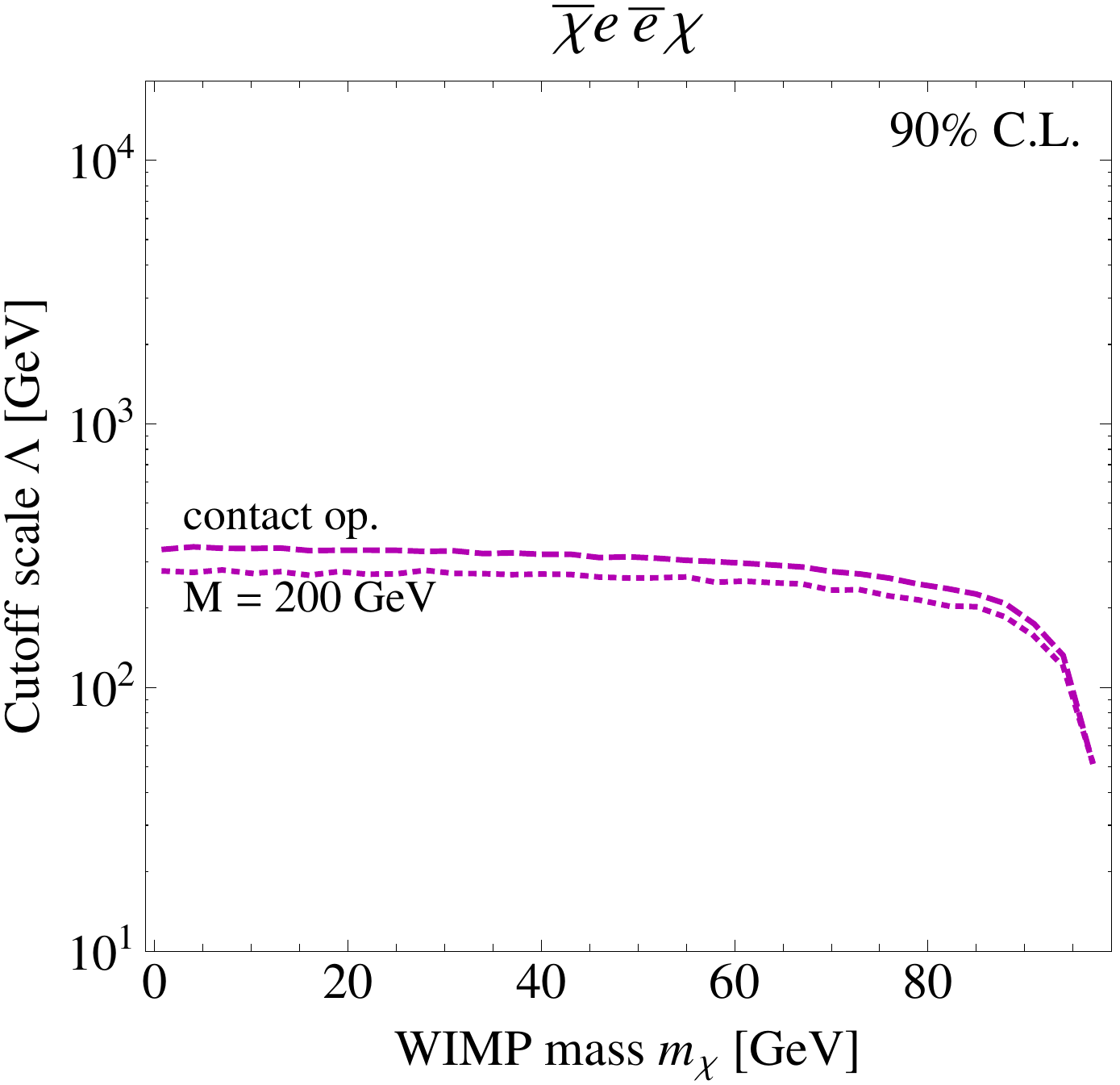}
  \end{center}
  \caption{DELPHI lower limits on the cutoff scale $\Lambda = M/\sqrt{g_e g_\chi}$ of
    the dark matter effective theory. Dashed lines have been computed under the
    assumption that the effective theory is valid up to LEP energies, whereas
    the dotted and solid lines are for cases where the mediator mass $M$ is so
    small that the effective theory breaks down. Once the mediator can be
    produced on-shell, its width $\Gamma$ becomes relevant, as demonstrated by
    the shaded regions.  $\Gamma_{\rm min}$ is the minimum allowed width of the
    mediator, where $g_e \approx g_\chi=M/\Lambda$, and $\Gamma_{\rm min}\gtrsim
    10^{-4}\,\gev$.  }
  \label{fig:cutofflight}
\end{figure}

Possible renormalizable theories that couple dark matter with the standard model fall into two general categories, which we will dub ``$s$-channel'' and ``$t$-channel'' mediators. In the first case the mediator is a neutral boson which has coupling vertices to $e^+e^-$ and to dark matter pairs. In this case the mediator may be almost arbitrarily light if its couplings with matter are sufficiently feeble. Of the operators we consider here, $s$-channel mediators give rise to operators of the form of (\ref{O1})--(\ref{O2}) at low energies.
In the second case dark matter is produced at colliders via a $t$-channel diagram, exchanging a charged mediator. The canonical example is supersymmetry where neutralino dark matter may be produced at LEP by exchanging a scalar selectron. At low energies this gives rise to the operator (\ref{O3}). Since the mediator is charged in this case, its mass should exceed about 110 GeV to evade direct LEP bounds.

In cases where the momentum flowing through the mediator in collider environments is of order the mediator mass $M$, the momentum-dependence of the propagator has to be taken into account.
In particular,  the amplitude will be proportional to 
\begin{equation}
\mathcal{A}\propto g_e\, g_\chi \frac{1}{q^2 - M^2+i M\, \Gamma}~,
\label{eq:amplitude}
\end{equation}
where $q$ is the 4-momentum carried by the mediator, $g_e$ ($g_\chi$) is the coupling of the mediator to electrons (dark matter) and $\Gamma$ is the total width of the mediator. In the case of an $s$-channel mediator $q^2 = s - 2 \sqrt{s} E_\gamma$ is \emph{positive}, while in the $t$-channel case $q^2$ is \emph{negative} and depends on the relative momentum between the two dark matter particles.  In the previous sections, where the massive mediator could effectively be integrated out, the higher dimension operators were suppressed by a scale $\Lambda$.  For the light mediator the LEP constraints become bounds on the geometric mean of the couplings $g_e$ and $g_\chi$, but for ease of comparison we can still formally define
\begin{align}
  \Lambda \equiv \frac{M}{\sqrt{g_e\,g_\chi}}~,
\end{align}  
and quote the bounds in terms of that quantity.

\begin{figure}[th]
  \begin{center}
    \includegraphics[width=8cm]{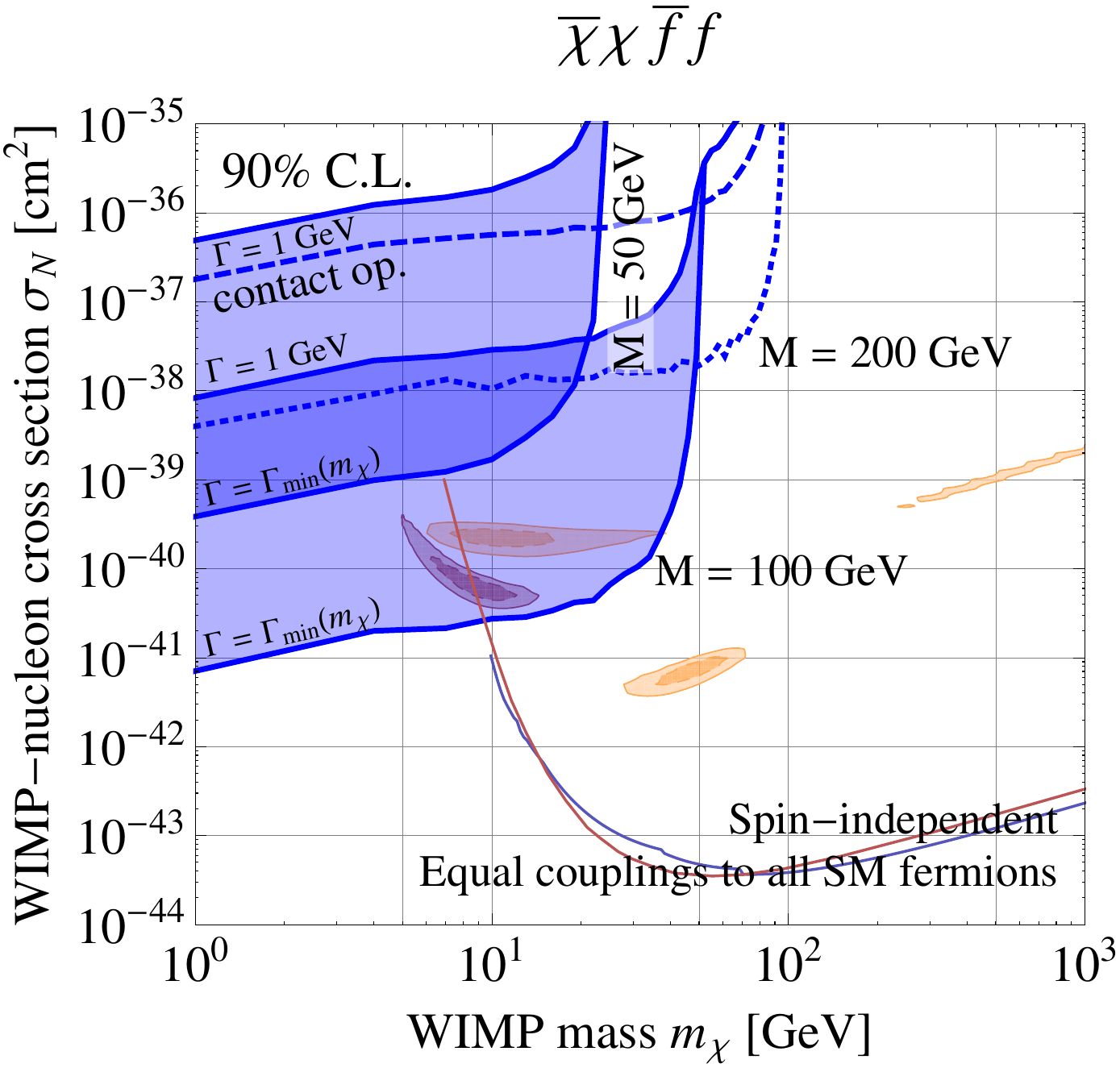}
    \includegraphics[width=8cm]{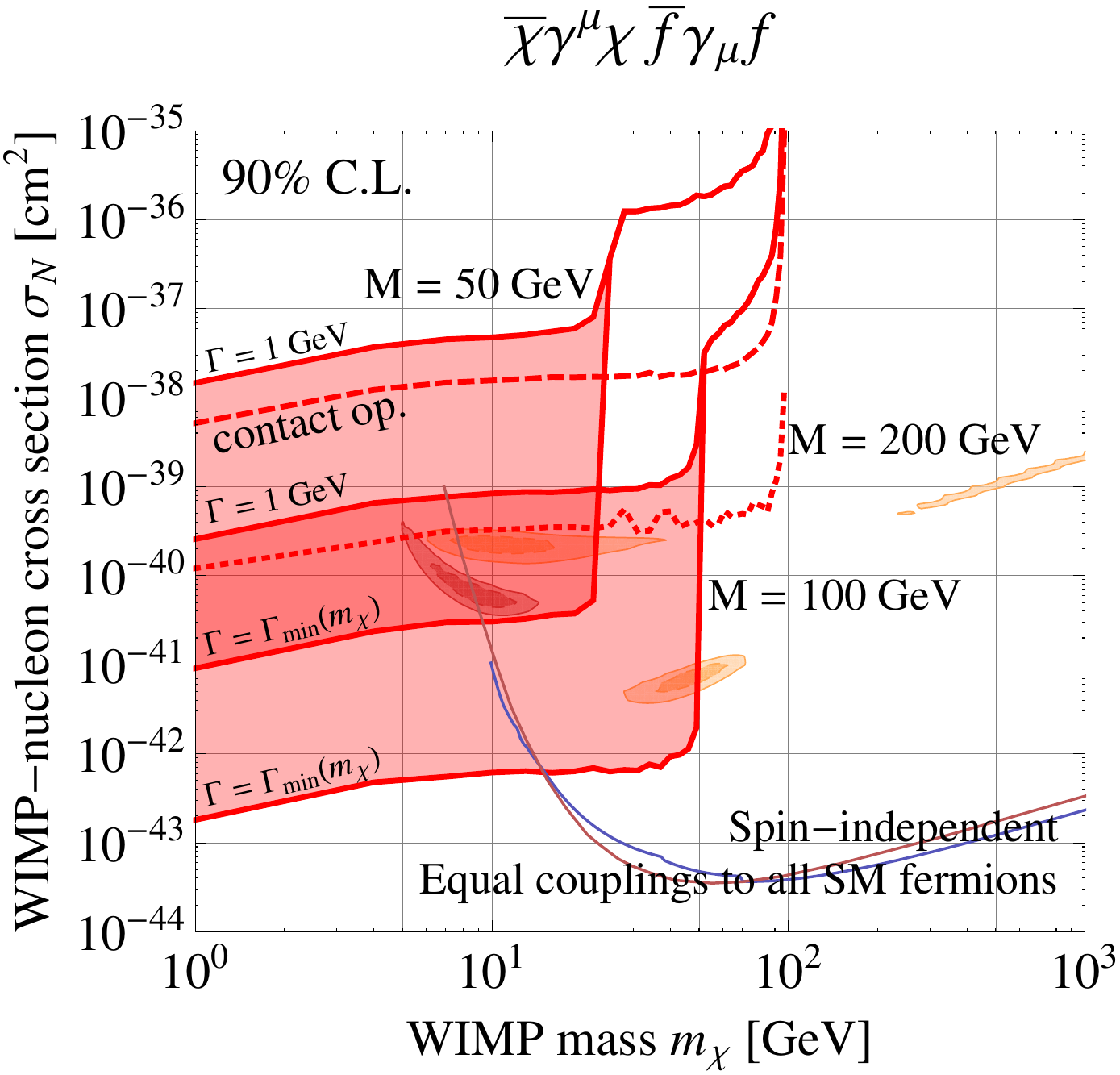} \\[0.2cm]
    \includegraphics[width=8cm]{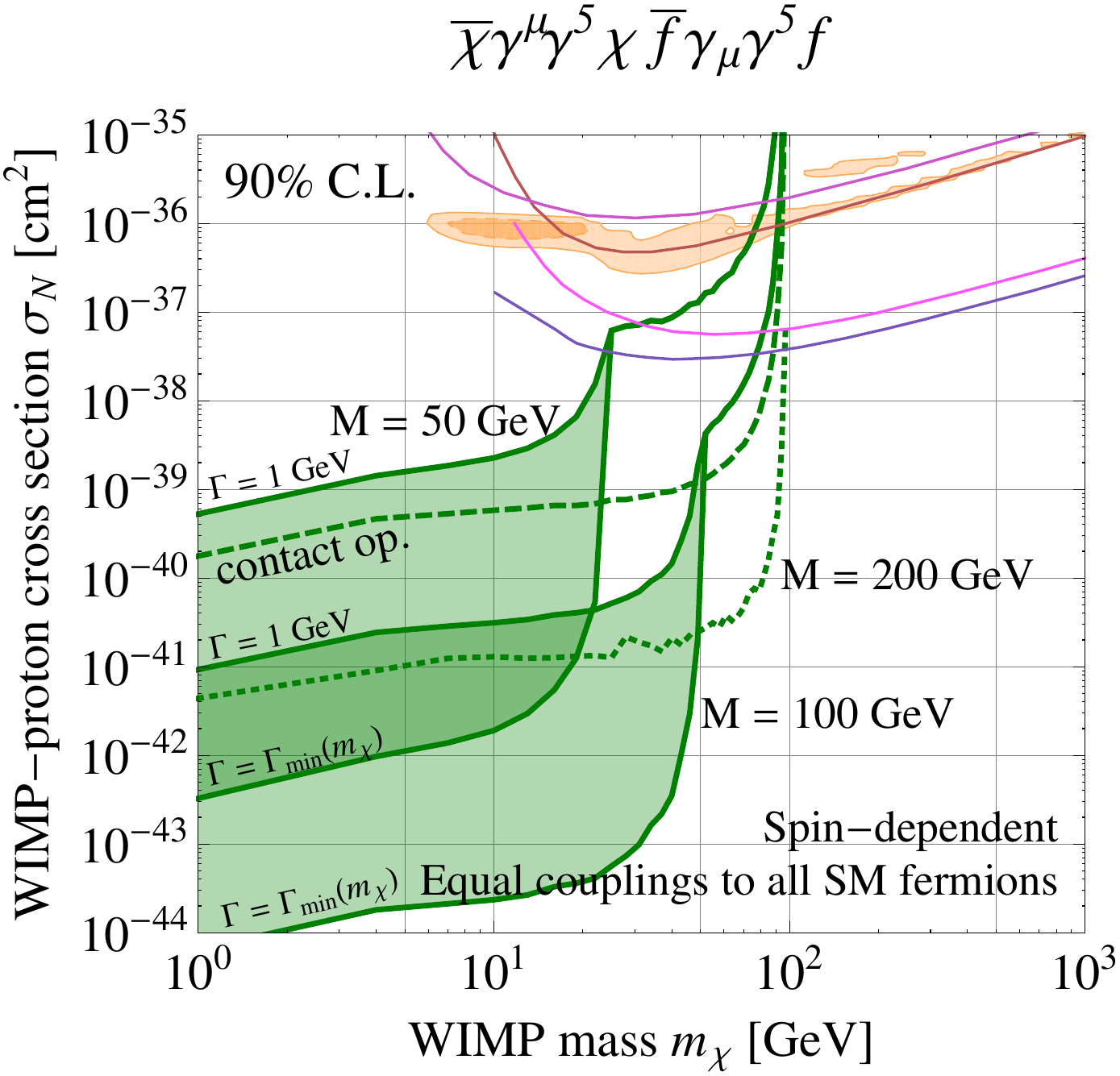}
    \includegraphics[width=8cm]{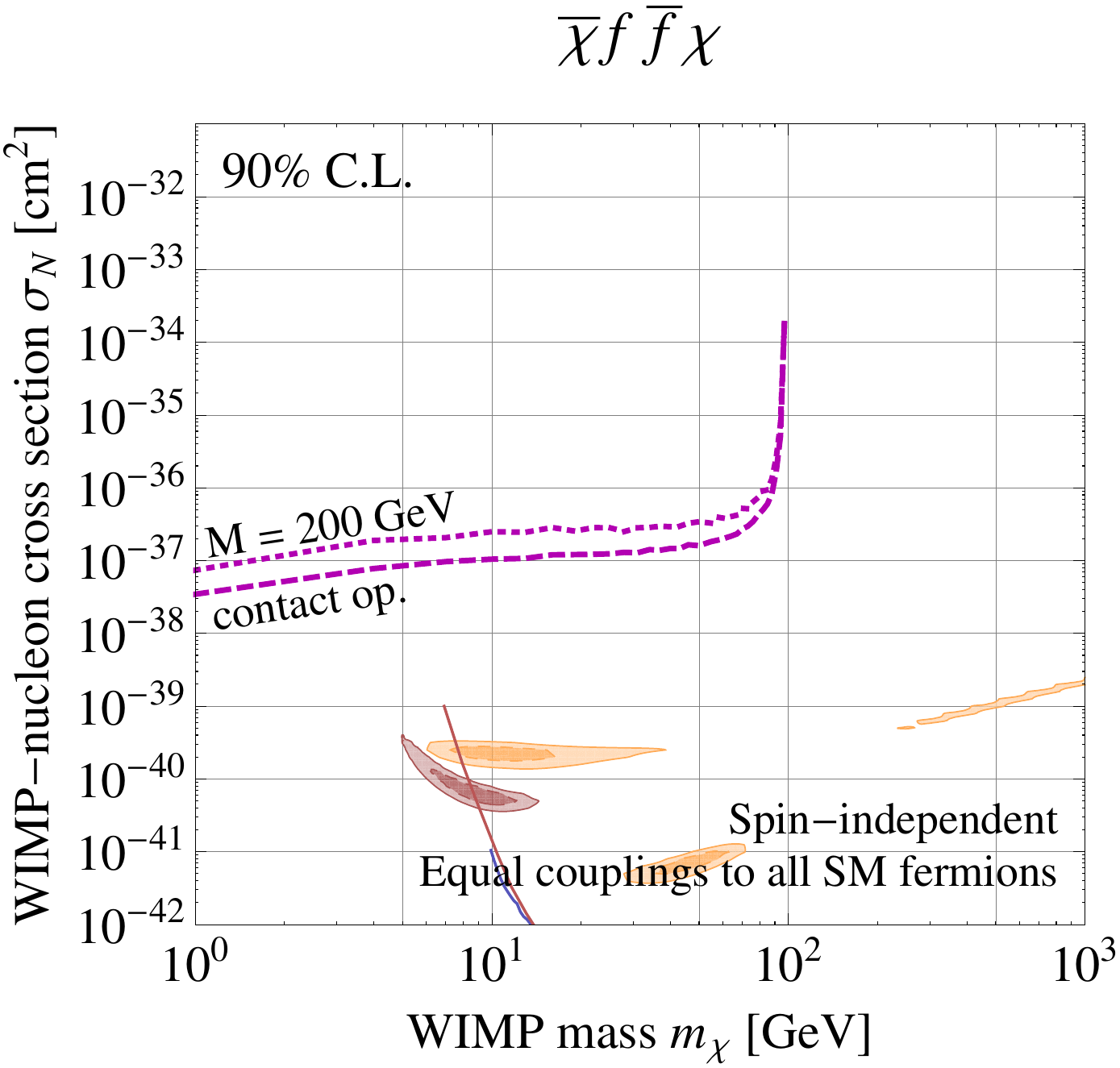}
  \end{center}
  \caption{DELPHI lower limits on the cross section for dark matter-nucleon scattering
    for different dark matter interaction models. As in Figure~\ref{fig:cutofflight}, from
    which the limits are derived, dashed lines correspond to a contact operator
    interaction between dark matter and electrons at LEP, while the solid and dotted
    lines are for interactions mediated by light particles. In the background,
    we show the constraints from the direct detection experiments XENON-100,
    CDMS, DAMA, and CoGeNT (upper left, upper right and lower right panels) and from
    DAMA, PICASSO, XENON-10, COUPP and SIMPLE (lower left panel), see
    fig.~\ref{fig:quarkcoupled} for details.}
  \label{fig:si-light}
\end{figure}

\begin{figure}[th]
  \begin{center}
    \includegraphics[width=8cm]{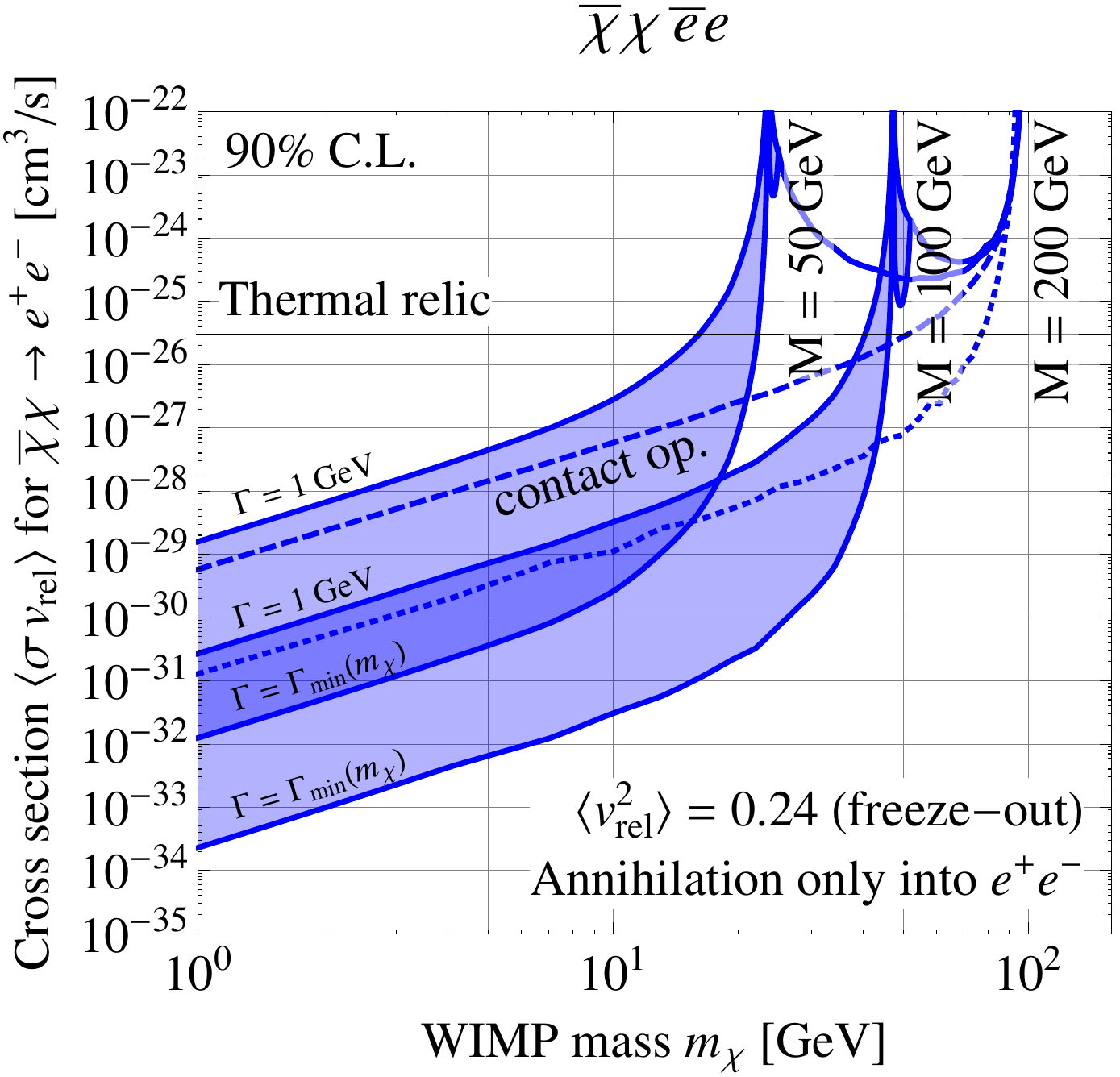}
    \includegraphics[width=8cm]{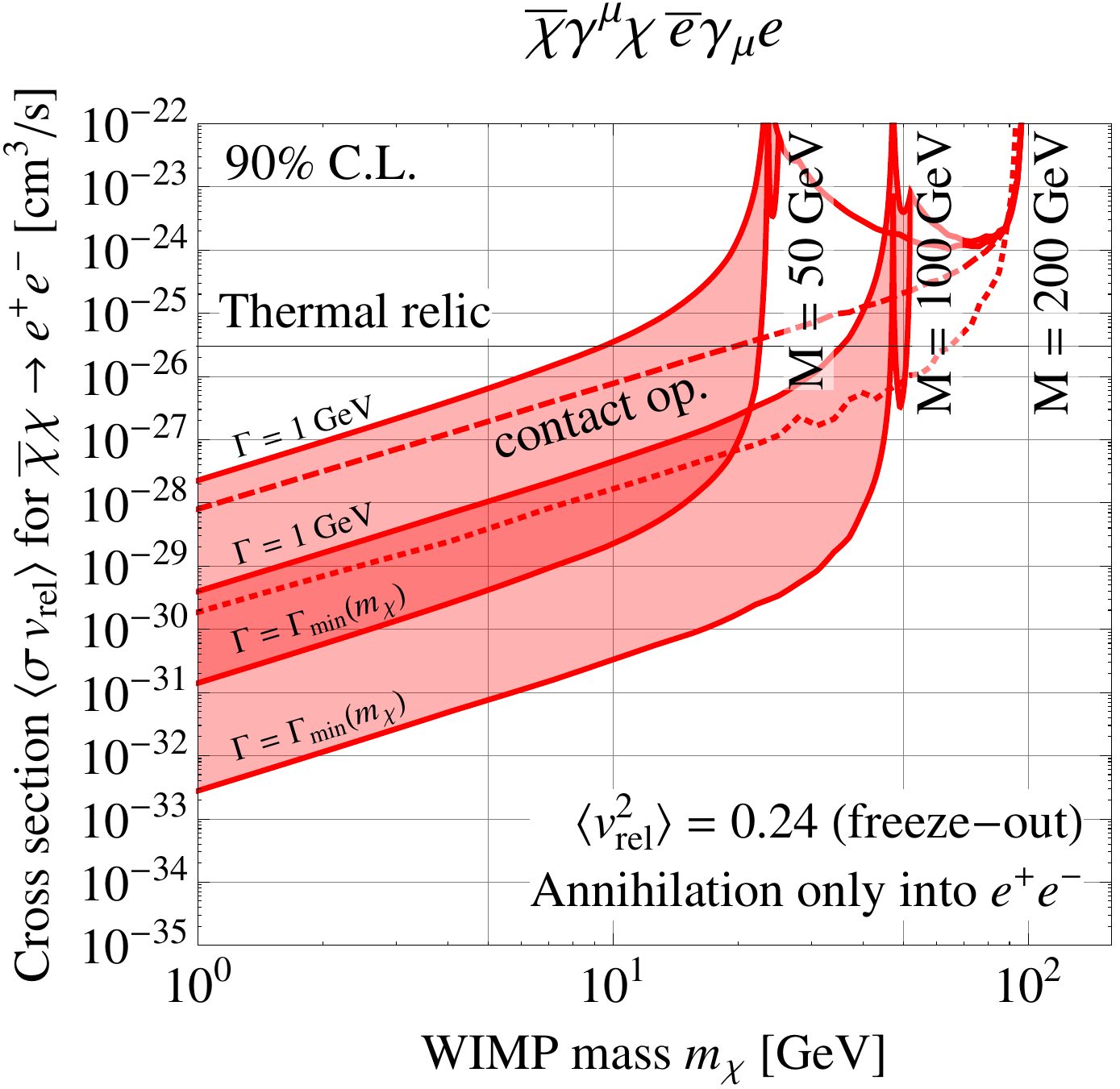} \\[0.2cm]
    \includegraphics[width=8cm]{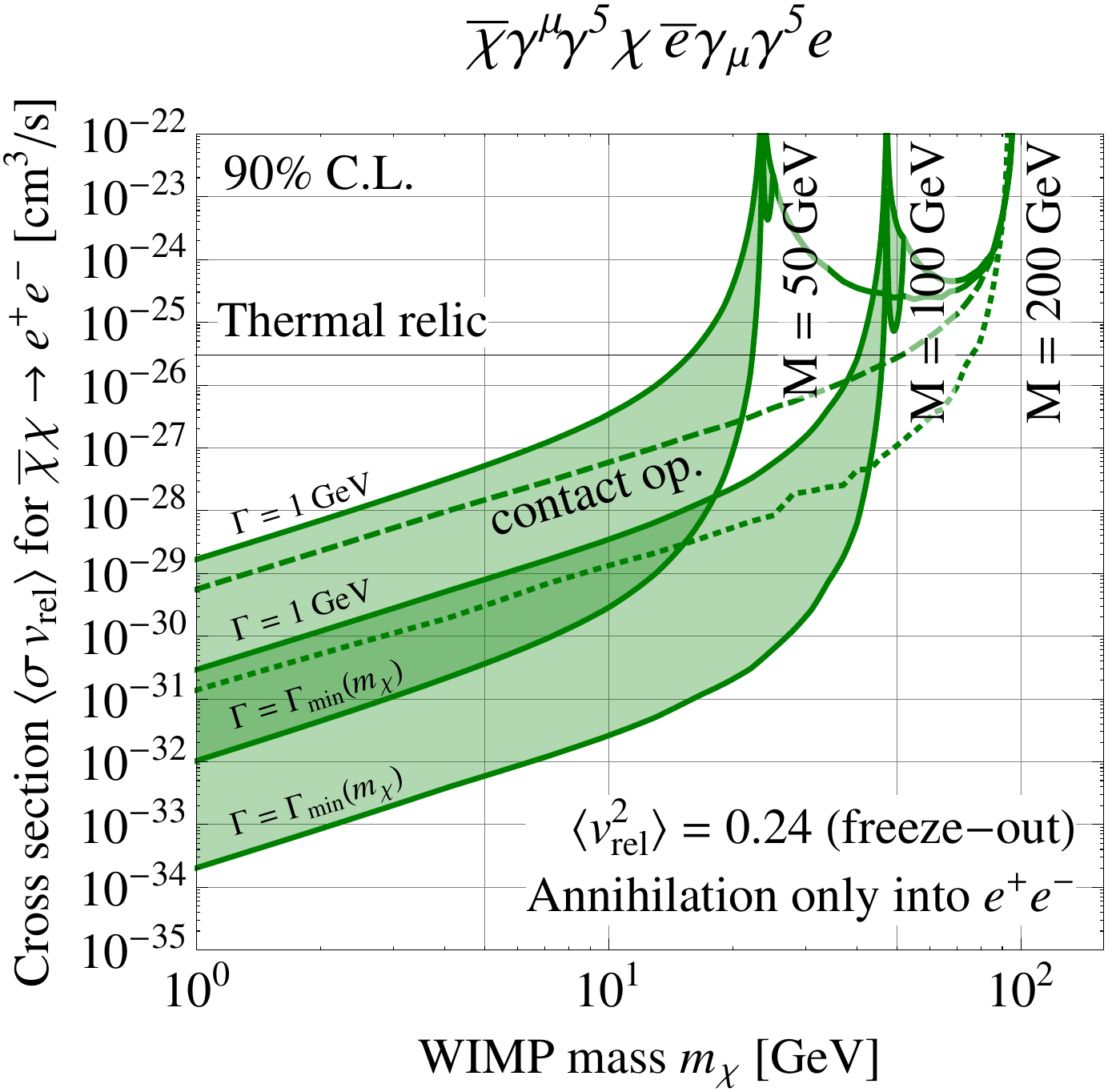}
     \includegraphics[width=8cm]{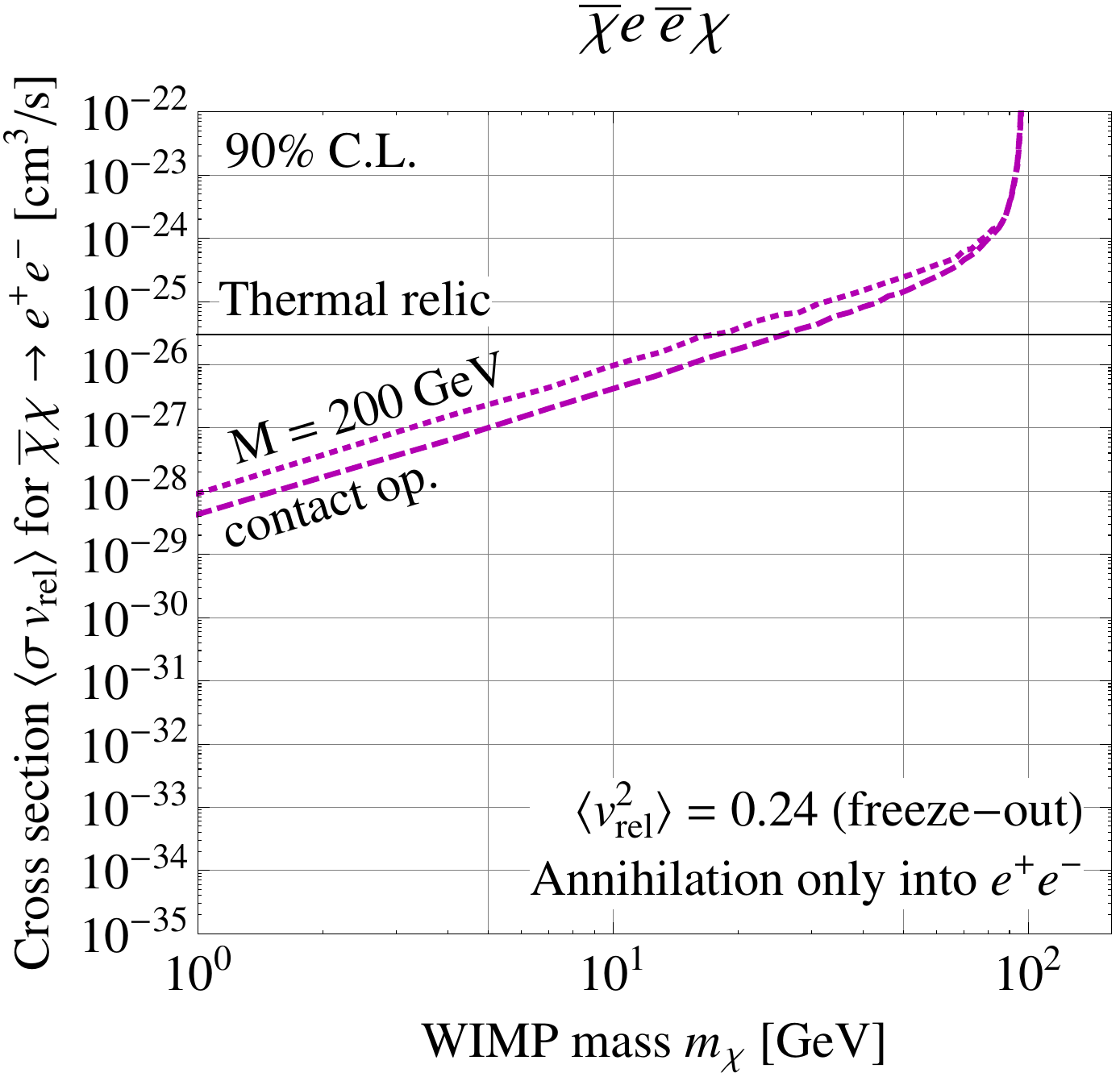}
  \end{center}
  \caption{LEP upper limits on the dark matter annihilation cross section
      $\ev{\sigma v}$ for different assumptions on the mass of the particle
      that mediates dark matter production and annihilation. We show limits
      only for the annihilation channel $\bar\chi \chi \to e^+ e^-$, which is
      the only one that can be probed model-independently at LEP. If dark
      matter has several annihilation channels, these limits can be
      straightforwardly (but in a model-dependent way) translated into limits
      on the total annihilation cross section, as done in the upper right and
      bottom panels of Figure~\ref{fig:annihilation}. As in Figure~\ref{fig:cutofflight}, from
      which the limits are derived, dashed lines correspond to a contact operator
      interaction between dark matter and electrons at LEP, while the solid and dotted
      lines are for interactions mediated by light particles.}
  \label{fig:annihilation-light}
\end{figure}

If the mediator and the dark matter are light enough to be produced on-shell at LEP, the bounds become sensitive to the width of the mediator $\Gamma$.  $\Gamma$ in turn depends on $g_e$, $g_\chi$ (and possibly on the couplings to other particles). Here, we will treat $\Gamma$ as a free parameter, but we note that, for any given value of $\sqrt{g_e\,g_\chi}$ (the quantity constrained by LEP), we can derive a lower limit $\Gamma_{\rm min}$ on $\Gamma$ by noting that
\begin{align}
  \Gamma = \frac{g_\chi^2}{24 \pi}M \sqrt{1 - \frac{4 m^2}{M^2}} (1+2 \frac{m^2}{M^2})
         + \frac{g_e^2}{24 \pi} M + \dots
  \label{eq:Gamma}
\end{align}
The first term comes from decay into dark matter, the second one from decay into electrons, and `\dots' stands for possible additional decay modes.  For fixed $\sqrt{g_e\,g_\chi}$ the width is minimized if $g_e \approx g_\chi$, and if $e^+ e^-$ and $\bar{\chi} \chi$ are the only allowed decay modes. If the latter assumption is true we can also place an upper bound on $\Gamma$ by setting $g_e = 4\pi$ and $g_\chi = M^2 / (g_e \Lambda_{\rm lim}^2)$ in \eqref{eq:Gamma}, where $\Lambda_{\rm lim}$ is minimum value of $\Lambda$ allowed by LEP.  In what follows we will take the mediator's width to be a free parameter and will consider the effects of $\Gamma_{\rm min} \le \Gamma \le 1\ \gev$. For dark matter coupling through a $t$-channel mediator, no resonant enhancement is possible, so the value of $\Gamma$ is irrelevant in this case.

The limits on $\Lambda = M/\sqrt{g_e g_\chi}$ for various interactions are presented in Figure~\ref{fig:cutofflight}. From (\ref{eq:amplitude}), we can understand the behavior of the dashed and dotted lines in this figure.  Consider first the $s$-channel case for $m_\chi > M/2$: There is no possibility of resonant production, so the mediator width is unimportant.  Comparing the cross sections for dark matter production at LEP in the contact operator and light mediator cases, we obtain 
\begin{align}
  \frac{d\sigma}{dE_\gamma} \Big|_{\text{light mediator}}
  = \frac{M^4}{(q^2 - M^2)^2} \, \frac{d\sigma}{dE_\gamma} \Big|_{\text{contact op.}} \,.
  \label{eq:xsec-light-nores}
\end{align}
with $q^2 = s - 2 \sqrt{s} E_\gamma$. Thus, for $M$ slightly below $\sqrt{s}$, there is partial cancellation between the $q^2$ and $M^2$ terms in the denominator, leading to an enhanced cross section and an improvement in the limit on $\Lambda$ compared to the contact operator case. For even smaller $M$, this cancellation is smaller and we expect the bound on $\Lambda$ to scale with $M$. This is confirmed by Figure~\ref{fig:cutofflight}.

On the other hand, if $2 m_\chi < M < \sqrt{s}$, the process $e^+ e^- \to \gamma \bar{\chi} \chi$ can proceed through an on-shell mediator, which leads to a peak in the monophoton spectrum reflecting the kinematics of a $2 \to 2$ scattering process. The absence of a strong peak in the DELPHI data, apart from standard model $Z$ production, places a strong constraint on this scenario. The constraint depends sensitively on the width of the mediator and scales as
\begin{align}
  \Lambda \propto 1 / \Gamma^{1/4} \,.
  \label{eq:Lambda-scaling}
\end{align}
This can be understood if we note that the resonant cross section for production of the mediator together with a single photon contains a factor
\begin{align}
  \frac{1}{\Lambda^4} \frac{1}{(E_\gamma - E_{\rm res})^2 + \Gamma^2 / 4} \,,
  \label{eq:xsec-light-res}
\end{align}
where $E_{\rm res}$ is the energy of the peak in the monophoton spectrum. Integrating \eqref{eq:xsec-light-res} over the photon energy $E_\gamma$, we find that the total cross section for on-shell production of the mediator is proportional to $1 / \Lambda^4 \Gamma$ (times factors that do not depend on $\Lambda$ or $\Gamma$), which explains equation~\eqref{eq:Lambda-scaling}. We have also confirmed the scaling of the bound on $\Lambda$ with $\Gamma^{-1/4}$ numerically.

Going back to Figure~\ref{fig:cutofflight} and comparing the limits on $\Lambda$ obtained for different types of operators---scalar, vector, and axial vector---we find that they are all comparable. The $t$-channel case is similar to the case of the $s$-channel away from resonance, except that the negative $q^2$ causes the denominator in equation~\eqref{eq:xsec-light-nores} to be always larger than the numerator, meaning that the bound on $\Lambda$ is always weaker in the light mediator case than for the contact operator. Furthermore, in the $t$-channel case there is obviously no on shell production of a mediator at low dark matter mass.

Even though the effective theory is not appropriate to describe production at LEP, it is still a good description of dark matter-nucleus scattering in direct detection experiments, where the exchanged momentum is very low. The procedure of translating our bound into direct detection limits is identical to that of Section~\ref{sec:dd}. We present these bounds in Figure~\ref{fig:si-light}, making the assumption that dark matter has equal couplings to all standard model quarks and leptons.

For non-resonant dark matter production ($s$-channel with $2m_\chi>M$ or $t$-channel), the presence of the light mediator in general severely weakens the LEP bounds on the direct detection cross section. As discussed below equation~\eqref{eq:xsec-light-nores}, however, there is a window of mediator masses where the bounds are marginally improved compared to the contact operator case. If the mediator can be produced on-shell and is sufficiently narrow, the bounds on direct detection rates are strengthened considerably.  In this case, the LEP constraints cover the (low mass) DAMA and CoGeNT-favored regions; for the vector operator a narrow resonance even impacts the DAMA region around $m_\chi\sim 50\,\gev$. However, we should emphasize again that these conclusions can be evaded if the coupling of dark matter to electrons is much smaller than its coupling to quarks.

Finally, we carry out a similar analysis to Section~\ref{sec:id} and compute the annihilation rate in the early universe in the case of a light mediator.  We consider only the case where the mediator couples exclusively to electrons and dark matter. Figure~\ref{fig:annihilation-light} shows that the LEP constraints on dark matter annihilation in the early universe change significantly if the mediator is light. The sharp peaks that occur at $m_\chi (1+\langle v^2\rangle/2) \approx M/2$ are due to resonant annihilation of dark matter, and the dips observed just above the peaks are due to the fact that resonant annihilation and the on-set of resonant production at LEP occur at slightly different values of $m_\chi$. As in Figures~\ref{fig:cutofflight} and \ref{fig:si-light},
the width of the mediator is of crucial importance for $m_\chi < M/2$.

\section{Conclusions}
\label{sec:conclusion}

Very little is known about the dark sector of particle physics.  It is usually assumed that dark matter couples, to varying degrees, to all fermions in the standard model, and strong constraints have been placed on its coupling to quarks by direct and indirect detection experiments and by the Tevatron. However, it is possible that dark matter has no coupling to quarks or at least that couplings to leptons are dominant.  In such a scenario, dark matter may be efficiently produced in collisions of electrons and positrons at LEP.  Irrespective of whether dark matter is leptophilic or not, LEP is an additional probe of its properties, and in this paper we have studied what LEP can say about the dark sector. Unlike dedicated dark matter searches in direct and indirect detection experiments, our LEP bounds do not suffer from astrophysical or atomic uncertainties.

One mode in which dark matter may be searched for at LEP, with relatively little model dependence, is its pair production in association with a hard photon.  The LEP experiments have searched for anomalous mono-photon events in their data sets, but have found no discrepancy from the prediction of the standard model.  Unlike at hadronic machines, at LEP the kinematics of the event can be completely determined allowing the standard model backgrounds to be more easily distinguished from dark matter production.  We used the mono-photon spectrum from the DELPHI experiment to place bounds upon the properties of dark matter that couples to electrons, see Figure~\ref{fig:cutoff}. In the first part of the paper, we worked in an effective theory framework, in which dark matter interactions are described by four-fermion contact operators, and we derived constraints on the suppression scale, $\Lambda$, of these operators.

We applied the LEP bounds on electron-dark matter coupling to constrain both the direct detection cross section and the annihilation rate of dark matter.  We considered both the case where dark matter couples equally to all leptons and a scenario in which dark matter couples equally to all standard model fermions.  Not surprisingly, for the ``leptophilic" scenario, where LEP is probing tree-level interactions but direct detection proceeds through a loop process, LEP bounds are highly competitive.  In fact, the bounds presented here rule out the DAMA favored region, excluding leptophilic dark matter as an explanation of the DAMA modulated events or the CoGeNT excess, see Figure~\ref{fig:leptononly}. 

In the case of equal couplings to quarks and charged leptons, the LEP bounds are complementary to direct detection bounds on spin-independent dark matter, see Figure~\ref{fig:quarkcoupled}.  They are weaker than existing direct detection bounds for dark matter mass $m_\chi$ larger than $\sim 4\,\gev$, but for light dark matter, $m_\chi\lesssim 4\,\gev$, they are significantly stronger.  For spin-dependent interactions, where direct detection constraints are relatively weak, LEP outperforms all other experiments up to its kinematic limit, $m_\chi\lesssim 80\,\gev$. LEP bounds are slightly stronger than those derived in \cite{Bai:2010hh} from Tevatron mono-jet searches, but do not extend to as high masses, and they depend on the assumption that dark matter has universal couplings to quarks and leptons. 

We have also used LEP bounds to constrain dark matter annihilation rates, both in the early universe and in present-day galaxies.  Below the LEP kinematic limit the LEP constraints are highly competitive. In particular, for $m_\chi \lesssim 80\,\gev$, they are stronger than those coming from Fermi-LAT observations of dwarf galaxies and of the galactic center, see Figure~\ref{fig:annihilation}.  They also provide a non-trivial constraint on a model invoked recently to explain a possible $\gamma$-ray excess at the galactic center~\cite{Hooper:2010mq}.

In the second part of the paper, we have repeated our analysis for the case where the interaction between dark matter and electrons cannot be treated as a contact operator.  We have ``UV completed'' the theory by introducing a particle that mediates dark matter-standard model interactions and have investigated LEP constraints as a function of the mediator mass and width. We find that, as long as dark matter cannot be produced through an on-shell mediator at LEP, our constraints are generally weaker than in the contact operator case (except for a narrow range of mediator masses close to the kinematic threshold of on-shell production). If the mediator mass $M$ is below the LEP center of mass energy, but larger than $2 m_\chi$, dark matter can be produced resonantly. In this case, the LEP constraint depends strongly on the width $\Gamma$ of the mediator---a model-dependent quantity---but if $\Gamma$ is small enough, the LEP constraint on the dark matter-electron coupling can be significantly stronger than for the contact operator case, see Figures~\ref{fig:cutofflight}, \ref{fig:si-light}, \ref{fig:annihilation-light}.

As the hunt for dark matter continues and we probe the dark sector on several fronts, both indirectly, directly and at the Tevatron and the LHC it is amusing to discover that there are non-trivial constraints still to be found in now completed experiments.  It seems that dark matter requires us to be students of history as well as physics.

\section*{Acknowledgments}
We thank Yang Bai, Graham Kribs, Andreas Kronfeld, and Zoltan Ligeti for discussions, and Dan Hooper for comments on a draft of the manuscript. 
PF thanks the Institute for Advanced Study for kind hospitality while this work was being completed and TRIUMF for providing a place to work while the wheels of the INS slowly turned. RH thanks the Berkeley Center for Theoretical Physics as well as the Stanford Institute for Theoretical Physics for their warm hospitality as this work was completed. YT is supported by a Fermilab Fellowship in Theoretical Physics. Fermilab is operated by Fermi Research Alliance, LLC, under Contract DE-AC02-07CH11359 with the United States Department of Energy.

\appendix
\section{Derivation of the average dark matter velocity in a dwarf galaxy}
\label{sec:vsq}

In this appendix, we discuss the derivation of the average dark matter velocity in
the Draco dwarf galaxy. We assume the radial distribution of dark matter in Draco
to follow a Navarro-Frenk-White (NFW) profile~\cite{Navarro:1995iw},
\begin{align}
  \rho(r) = \rho_s \frac{r_s}{r} \frac{r_s^2}{(r + r_s)^2} \,,
\end{align}
with scale radius $r_s = 2.09$~kpc and scale density $\rho_s =
0.98$~GeV/cm$^3$~\cite{Abdo:2010ex}. 

We then use the Eddington formula~\cite{Binney:1950} 
\begin{align}
  f(r, v) &= \frac{1}{\sqrt{8} \pi^2} \frac{d}{d\mathcal{E}}
    \int_0^\mathcal{E} \frac{d\rho}{d\Psi} \frac{d\Psi}{\sqrt{\mathcal{E} - \Psi}} \\
  &= \frac{1}{\sqrt{8} \pi^2} \int_0^\mathcal{E} \! d\Psi \frac{1}{\sqrt{\mathcal{E} - \Psi}}
     \frac{d^2\rho}{d\Psi^2} + \frac{1}{\sqrt{\mathcal{E}}}\frac{d\rho}{d\Psi}\bigg|_{\Psi = 0}
  \label{eq:Eddington}
\end{align}
to  translate $\rho(r)$ into the velocity distribution $f(r, v)$ at
radius $r$. Here, $\Psi(r) = -G \int_R^\infty \!dr \, M(r) / r^2$ is (minus) the gravitational
potential at radius $r$, which is determined by the enclosed mass
$M(r) = \int_0^r \! dr\, 4\pi r^2 \, \rho(r)$, and $\mathcal{E}(r, v) =
\Psi(r) - \frac{1}{2} v^2$ is (minus) the dark matter energy per unit mass. The dark matter density $\rho$ is treated as a function of $\Psi$ rather than $r$ here, which is well-defined if
$\Psi(r)$ is a monotonic function of $r$; the NFW profile has this property.
The resulting velocity distribution $f(r, v)$ satisfies
the normalization condition
\begin{align}
   \rho(r)=4\pi\int\!dv \,  v^2 f(r, v) \,.
\end{align}
The annihilation rate of dark matter is proportional to $\rho^2 \sigma v$, thus to obtain $\ev{v^2}$, the average dark matter velocity in the Draco
dwarf galaxy quoted in Section~\ref{sec:id}, we compute
\begin{align}
  \ev{v^2} = \frac{1}{N} \int_0^\infty\!dr \, 4\pi r^2 \rho^2(r)
                   \int_0^1\!dv \, 4\pi v^4 f(r,v)
\end{align}
with the normalization constant $N = \int_0^\infty\!dr \, 4\pi r^2 \rho^2(r)$.  We find $\ev{v^2}\approx (34.7\ \mathrm{km}/\mathrm{s})^2$, which corresponds to $\ev{v_{\rm rel}^2}\approx (69.3\ \mathrm{km}/\mathrm{s})^2$.

\bibliographystyle{apsrev}
\bibliography{./lep}

\end{document}